\documentclass[letterpaper, 12pt]{article}
\usepackage{amsmath}
\usepackage{amssymb}

\title{Absorption and quasinormal modes of classical fields propagating on $3D$ and $4D$ de Sitter spacetime}

\author{A.\ L\'opez-Ortega\thanks{Electronic address: alfredo@fis.cinvestav.mx} \\Departamento de F\'{\i}sica CINVESTAV IPN\\ Apdo. Postal 14-740, 07000 M\'exico D. F., M\'exico.  }

\begin{document}

\maketitle

\begin{abstract}

We extensively study the exact solutions of the massless Dirac equation in $3D$ de Sitter spacetime that we published recently. Using the Newman-Penrose formalism, we find exact solutions of the equations of motion for the massless classical fields of spin $s=\tfrac{1}{2},1,2$ and to the massive Dirac equation in $4D$ de Sitter metric. Employing these solutions, we analyze the absorption by the cosmological horizon and de Sitter quasinormal modes. We also comment on the results given by other authors.\\



\end{abstract}

\newpage

\section{Introduction}
\label{csix:Introduction}

The properties of de Sitter background has recently been studied \cite{Spradlin:2001pw}. In particular, it was found that this spacetime has several applications in different topics of theoretical physics such as the dS-CFT correspondence, inflation, quantum fields in curved spacetime, thermodynamics of horizons \cite{Park:1998qk}--\cite{Riess:1998cb}, since the simplicity of its metric allows us to investigate in detail (in some cases in exact form) many physical phenomena. Understanding of these phenomena in de Sitter background will be useful in more complex spacetimes, which have more relevance from a physical point of view. 

In order to analyze the properties of a spacetime, a method often used consists in studying the propagation of a classical field. In the present paper we continue to investigate the dynamics of different classical fields propagating in de Sitter metric as in Refs.\ \cite{Myung:2003cn}--\cite{Otchik:1985ih}. First, we find exact solutions of the equations of motion for several classical fields moving in de Sitter spacetime. In particular, for the massive Dirac equation we calculate a more simple set of solutions than those found in Ref.\ \cite{Otchik:1985ih}. Using these solutions we study the absorption of the cosmological horizon and extend some of the results given in Refs.\ \cite{Myung:2003cn, Myung:2003ki}. Following these studies we compute the radial flux of a classical field propagating in de Sitter background when its frequency is a real number and its radial function is regular at $r=0$. Moreover, we briefly discuss the implications of these results and we enumerate the differences among our results and those of other references.

We expect that the quasinormal modes (QNMs) of the black holes will be a part of the signal that the gravitational wave detectors will detect in the near future \cite{Kokkotas:1999bd}. More recently, owing to the dS-CFT and AdS-CFT correspondences \cite{Strominger:2001pn, Bousso:2001mw}, \cite{Maldacena:1997re}, there is an increased interest in calculating the values of the quasinormal (QNM) frequencies in several asymptotically de Sitter  \cite{Abdalla:2002hg, Abdalla:2002rm} and anti-de Sitter black holes \cite{Horowitz:1999jd, Birmingham:2001pj}. Several papers have studied the QNMs of de Sitter spacetime \cite{Choudhury:2003wd}--\cite{Abdalla:2002hg}; nevertheless, these papers report different results on their existence.\footnote{The anti-de Sitter QNMs has also been calculated in Ref.\ \cite{Natario:2004jd}} 

Motivated by this discussion in the literature on the cases in which the de Sitter QNM frequencies of massless fields are well defined and using the exact solutions that we previously found, in this work we calculate the de Sitter QNMs of a massless Klein Gordon field, an electromagnetic field, a gravitational perturbation and a massive Dirac field in $4D$ and of a massless Dirac field in $3D$. We also comment on some of the previous results.

This paper is organized as follows. In Section \ref{csix:dS3-Dirac-massless} we review some features of the dynamics of a massless Dirac field propagating in $3D$ de Sitter spacetime. We improve the results given in Sections 3 and 5 of Ref.\ \cite{Lopez-Ortega:2004cq} by using a different triad to that of the previously mentioned reference and we find additional exact solutions of the radial equations. In Section \ref{csix:dS4-massless-fields} we obtain exact solutions of the equations of motion for the massless fields of spin $s=\tfrac{1}{2}, 1, 2$ propagating in $4D$ de Sitter metric. In Section \ref{csix:dS4-massive-Dirac} we develop another method which allows us to find exact solutions of the massive Dirac equation in $4D$ de Sitter spacetime. Employing the new exact solutions, in each of the previous sections, we can calculate the absorption and QNM frequencies of some classical fields moving in de Sitter background. Finally, in Section \ref{csix:End-comments} we make some comments on these results. 

In Appendix \ref{appendix:half-integer-m} we review some facts used in Section \ref{csix:dS3-Dirac-massless}. In Appendix \ref{appendix:quasinormal-modes-4D} we present a different method for calculating the QNM frequencies given in Subsection \ref{csix:subsection:quasinormal-modes}. In Appendix \ref{appendix:solution-dirac-equation}, to compare with the solutions calculated in Sec.\ \ref{csix:dS4-massive-Dirac}, we briefly describe the transformations employed in Ref.\ \cite{Otchik:1985ih} to find exact solutions of the massive Dirac equation in $4D$ de Sitter spacetime.

\section{Massless Dirac field in $3D$ de Sitter spacetime }
\label{csix:dS3-Dirac-massless}

In Ref.\ \cite{Lopez-Ortega:2004cq} we have shown that using a null triad (Eq.\ (3) of Ref.\ \cite{Lopez-Ortega:2004cq}) it is possible to find exact solutions of the massless Dirac equation when the background spacetime is $3D$ de Sitter. In the present section, we obtain exact solutions of the massless Dirac equation by means of a slightly different method to that used in \cite{Lopez-Ortega:2004cq}. Employing these solutions we study the absorption and QNMs of a massless Dirac field when it is moving in $3D$ de Sitter metric.

Note that in Ref.\ \cite{Lopez-Ortega:2004cq} we do not consider all the cases for which it is possible to find exact solutions of the radial differential equations; therefore, in the following we also give other values for the parameters of the solutions that were not analyzed in this reference.

In the circularly symmetric metric 
\begin{equation} \label{metric}
{\rm d}s^2=P(r)^2 {\rm d}t^2 - Q(r)^2 {\rm d} r^2 -r^2 {\rm d} \theta^2,
\end{equation}
if we use the diagonal triad
\begin{eqnarray} \label{triad}
e_1^{\,\,\,\mu} &=&  \left( \frac{1}{P}, 0,0
 \right), \nonumber \\
e_2^{\,\,\,\mu} &=&  \left( 0,\frac{1}{Q}, 
0 \right), \\
e_3^{\,\,\,\mu} &=& \left(0,0,\frac{1}{r} \right), \nonumber
\end{eqnarray}
and the representation of the Gamma matrices $\gamma_{a}$,
\begin{eqnarray} \label{gamma}
\gamma_1 &=&  \sigma_1   = 
 \left( \begin{array}{cc} 0 & 1 \\ 1 & 0  \end{array} 
\right), \nonumber \\
\gamma_2 &=&  i \sigma_2  = 
\left( \begin{array}{cc} 0 & 1 \\- 1 & 0  \end{array} \right) 
, \\
\gamma_3 &=& i \sigma_3 = \left( \begin{array}{cc} i & 0 \\ 0 & -i  
\end{array} \right),  \nonumber
\end{eqnarray} 
where $\sigma_i$ denote the Pauli matrices; the massless Dirac equation 
\begin{equation} \label{Dirac equation}
\gamma^a \nabla_a \Psi = \gamma^{a} ( e_a + \Gamma_a ) \Psi = 0
\end{equation} 
becomes
\begin{equation} \label{Dirac matrix form}
\left( \begin{array}{cc} - \tfrac{i \partial_\theta }{r}  & \frac{\partial_t}{P} - \frac{\partial_r}{Q} - \frac{1}{2 Q} 
\left( \frac{1}{r} + \frac{P^\prime}{P} \right) \\  \frac{\partial_t}{P} + \frac{\partial_r}{Q} + 
\frac{1}{2 Q} \left( \frac{1}{r} + \frac{P^\prime}{P} \right) & \tfrac{i \partial_\theta}{r}   
\end{array} \right) \left( \begin{array}{c} \Psi_1 \\ \Psi_2 \end{array} 
\right) = 0.
\end{equation} 

We propose a separable solution of Eq.\ (\ref{Dirac matrix form}) in the form \cite{Lopez-Ortega:2004cq}
\begin{equation} \label{separable form}
\Psi = e^{- i \omega t} e^{im \theta} \left( \begin{array}{c} 
R_1^{(m)}(r) \\ R_2^{(m)}(r) \end{array} \right), 
\end{equation} 
where $m=\pm \tfrac{1}{2}, \pm \tfrac{3}{2}, \pm \tfrac{5}{2}, \dots$\footnote{See Appendix \ref{appendix:half-integer-m}.} Using this ansatz we find that the functions $R_1^{(m)}$ and $R_2^{(m)}$ satisfy the coupled differential equations\footnote{As in Ref.\ \cite{Lopez-Ortega:2004cq}, for simplicity in the notation, we only write the superscript $(m)$ of the radial functions when it is strictly necessary.}
\begin{eqnarray} \label{csix:R coupled}
\left\{ \frac{r}{Q} \frac{{\rm d}}{{\rm d}r} + \frac{i \omega r}{P} + 
\frac{r}{2 Q} \left( \frac{1}{r} + \frac{P^\prime}{P} \right) \right\} R_2 = 
m R_1 ,\nonumber \\
\left\{ \frac{r}{Q} \frac{{\rm d}}{{\rm d}r} - \frac{i \omega r}{P} + 
\frac{r}{2 Q} \left( \frac{1}{r} + \frac{P^\prime}{P} \right) \right\} R_1 = 
m R_2 .
\end{eqnarray}

These equations are equal to Eqs.\ (12) of Ref.\ \cite{Lopez-Ortega:2004cq}. To find exact solutions of the radial equations (\ref{csix:R coupled}) when the background spacetime is $3D$ de Sitter, the method employed in Ref.\ \cite{Lopez-Ortega:2004cq} directly uncoupled the functions $R_1$ and $R_2$ from the system of differential equations (\ref{csix:R coupled}). We simplified to hypergeometric differential equations the second order differential equations obtained for $R_1$ and $R_2$ by means of an ansatz and a change of variable.

To solve the coupled differential equations (\ref{csix:R coupled}) in $3D$ de Sitter metric, that is, when
\begin{equation} \label{csix:definition P}
P^2 = \frac{1}{Q^2} = 1 -\frac{r^2}{l^2},
\end{equation} 
we propose that the radial functions $R_1$ and $R_2$ take the form (see also Ref.\ \cite{Lopez-Ortega:2005ep})
\begin{eqnarray} \label{csix:ansatz-R}
R_1 = g \tilde{R}_1, \nonumber \\
R_2 = g \tilde{R}_2,
\end{eqnarray}
where 
\begin{equation} \label{csix:g-definition-R}
g = \frac{1}{z^{\frac{1}{2}} (1-z^2)^{\frac{1}{4}}},
\end{equation} 
and we define the variable $z$ by $r =z l$.

Substituting Eqs.\ (\ref{csix:ansatz-R}) and (\ref{csix:g-definition-R}) into Eqs.\ (\ref{csix:R coupled}), we get the simplified system of differential equations
\begin{eqnarray}
\left(\frac{{\rm d}}{{\rm d}z} + \frac{i \tilde{\omega}}{P^2} \right)\tilde{R}_2 = \frac{m}{P z} \tilde{R}_1, \nonumber \\
\left(\frac{{\rm d}}{{\rm d}z} - \frac{i \tilde{\omega}}{P^2} \right)\tilde{R}_1 = \frac{m}{P z} \tilde{R}_2,
\end{eqnarray}
where $\tilde{\omega} = \omega l$. From this system of differential equations it is possible to find that the function $\tilde{R}_1$ satisfies the differential equation
\begin{eqnarray} \label{csix:radial-decoupled}
\left[ \frac{{\rm d}^2}{{\rm d}z^2} + \left(\frac{1}{z} - \frac{z}{P^2}\right)\frac{{\rm d}}{{\rm d}z} + \frac{\tilde{\omega}^2}{P^4}- \frac{i \tilde{\omega} z}{P^4} -\frac{i\tilde{\omega}}{P^2 z}-\frac{m^2}{z^2P^2} \right]\tilde{R}_1 = 0.
\end{eqnarray} 
The function $\tilde{R}_2$ satisfies a similar equation.

Now, making the change of variable \cite{Lopez-Ortega:2004cq}, \cite{Suzuki:1995nh}
\begin{equation} \label{csix:y-definition}
y=\frac{1-z}{1+z}
\end{equation} 
we find that the differential equation (\ref{csix:radial-decoupled}) becomes
\begin{align} \label{csix:decoupled-y}
\left[ \frac{{\rm d}^2}{{\rm d}y^2} +\left(\frac{1}{2y}- \frac{1}{1-y} \right)\frac{{\rm d}}{{\rm d}y}  \right. + & \frac{\tilde{\omega}^2 - i \tilde{\omega}}{4y^2}  \nonumber \\
&\left.- \frac{ i \tilde{\omega}}{2 y (1 -y)} - \frac{m^2}{y(1-y)^2}\right]\tilde{R}_1 =0,
\end{align} 
($\tilde{R}_2$ satisfies a similar equation).

Making the following ansatz for $\tilde{R}_I$,
\begin{eqnarray} \label{csix:ansatz-R-y-radial}
\tilde{R}_I = (1-y)^{B_I}y^{C_I} \hat{R}_I,
\end{eqnarray} 
with $I=1,2$, and substituting into Eq.\ (\ref{csix:decoupled-y}) (and into the corresponding equation for $\tilde{R}_2$), we get that the functions $\hat{R}_I$ must be solutions of the differential equations\footnote{We notice that the symbols $A$, $B$, $C$, $a$, $b$, and $c$ represent different quantities in each section of this paper.}
\begin{equation} \label{csix:hipergeometric-y}
\left[y(1-y) \frac{{\rm d}^2}{{\rm d}y^2} + (c_I - (a_I+b_I+1)y)\frac{{\rm d}}{{\rm d}y} -a_I b_I  \right]\hat{R}_I = 0,
\end{equation} 
where
\begin{align}
&a_1 = C_1 + B_1 + \frac{1}{2}-\frac{i \tilde{\omega}}{2}, &a_2& = C_2 + B_2 + \frac{1}{2}+\frac{i \tilde{\omega}}{2}, \nonumber \\
&b_1 = C_1 + B_1 + \frac{i \tilde{\omega}}{2} , &b_2&=C_2 + B_2 - \frac{i \tilde{\omega}}{2},\nonumber \\
&c_1 = 2 C_1 + \frac{1}{2}, &c_2&= 2 C_2 + \frac{1}{2},
\end{align}
provided the parameters $B_1$, $B_2$, $C_1$, and $C_2$ take the values
\begin{align} \label{csix:B_I-C_I-two-values}
&B_1= \pm |m|, &B_2&= \pm |m|, \nonumber \\
&C_1=\frac{1}{4}\pm\frac{1}{4}\pm\frac{i \tilde{\omega}}{2}, &C_2&=\frac{1}{4}\mp\frac{1}{4}\pm\frac{i \tilde{\omega}}{2}.
\end{align} 

Note that in Ref.\ \cite{Lopez-Ortega:2004cq} we considered that the parameters $B_I$ and $C_I$ take only one value (Eq.\ (26) of Ref.\ \cite{Lopez-Ortega:2004cq}); however, we see in Eq.\  (\ref{csix:B_I-C_I-two-values}) that these quantities can take two values. As a consequence, we get a larger number of solutions of the radial differential equation (\ref{csix:radial-decoupled}). This also happens for the massive Klein Gordon equation in de Sitter metric \cite{Abdalla:2002hg} and for the Dirac equation in a 3$D$ black hole of EMD$\Lambda$ theory \cite{Lopez-Ortega:2005ep}.

In the rest of this section we use the variable $v$ defined by (see Ref.\ \cite{Lopez-Ortega:2004cq})
\begin{equation}
v = 1-y.
\end{equation} 
Therefore the differential equations (\ref{csix:hipergeometric-y}) become
\begin{equation} \label{csix:hipergeometric-v}
\left[v(1-v) \frac{{\rm d}^2}{{\rm d}v^2} + (\tilde{c}_I - (\tilde{a}_I+\tilde{b}_I+1)v)\frac{{\rm d}}{{\rm d}v} -\tilde{a}_I \tilde{b}_I  \right]\hat{R}_I = 0,
\end{equation} 
where \cite{b:DE-books}
\begin{align} \label{csix:abc-hipergeometric-v}
&\tilde{a}_1 = a_1, &\tilde{a}_2& = a_2, \nonumber \\
&\tilde{b}_1 = b_1, &\tilde{b}_2& = b_2, \nonumber \\
&\tilde{c}_1 = 2 B_1 + 1, &\tilde{c}_2& = 2 B_2 + 1.
\end{align} 

As is well known, the solutions of the differential equations (\ref{csix:hipergeometric-v}) are hypergeometric functions \cite{b:DE-books}. In the following, we study in detail the case $B_1= B_2 = |m|$. Later we shall comment what happens when $B_1$ and $B_2$ take other values.

From Eqs.\ (\ref{csix:abc-hipergeometric-v}) we note that $\tilde{c}_1$ and $\tilde{c}_2$ are positive integers. Thus the solutions of the differential equations (\ref{csix:hipergeometric-v}) are \cite{b:DE-books}
\begin{align} \label{csix:regular solution}
\hat{R}_{I}^{(1)} &= {}_{2}F_{1}(\tilde{a}_I,\tilde{b}_I;\tilde{c}_I;v), \\
\hat{R}_{I}^{(2)} &= {}_{2}F_{1}(\tilde{a}_I,\tilde{b}_I;\tilde{c}_I;v) \ln (v) \nonumber \\
& + \frac{(\tilde{c}_I - 1)!}{\Gamma(\tilde{a}_I) \Gamma(\tilde{b}_I)} \sum_{q=1}^{\tilde{c}_I - 1} (-1)^{q-1} (q - 1)! \frac{\Gamma(\tilde{a}_I - q) \Gamma(\tilde{b}_I -q)}{(\tilde{c}_I -q - 1)!} v^{-q} \nonumber \\
&  + \sum_{q=0}^{\infty} \frac{(\tilde{a}_I)_q (\tilde{b}_I)_q}{q! (\tilde{c}_I)_q} v^q \left[\psi(\tilde{a}_I + q) + \psi(\tilde{b}_I + q) - \psi(\tilde{c}_I+q) - \psi(1 + q) \right. \nonumber \\
  & \left. \hspace{0cm} - \psi(\tilde{a}_I -  \tilde{c}_I +1) - \psi(\tilde{b}_I - \tilde{c}_I + 1) + \psi(1) + \psi(\tilde{c}_I -1) \right], \label{eq:2-option}
\end{align} 
where $\psi(x) = {\rm d} \ln \Gamma (x) / {\rm d} x$.

Employing Eqs.\ (\ref{csix:ansatz-R}), (\ref{csix:ansatz-R-y-radial}), and (\ref{eq:2-option}) it is possible to show that the radial functions that include $\hat{R}_{I}^{(2)}$ are divergent at $r=0$ (the origin of the de Sitter spacetime). We have no reasons to expect singularities of the fields at the origin of the de Sitter metric; therefore, in the present work we study the solutions of the equations of motion that are regular at this point, as was the case in Refs.\ \cite{Bousso:2001mw}, \cite{Myung:2003cn}--\cite{Lopez-Ortega:2004cq}. Thus we only consider radial functions that include $\hat{R}_{I}^{(1)}$.

The parameters $C_1$ and $C_2$ can take two different values for given $B_1$ and $B_2$; then, it follows that it is necessary to study four cases for the possible permutations of $C_1$ and $C_2$, (something similar happens in the analysis made in Refs.\  \cite{Abdalla:2002hg}, \cite{Lopez-Ortega:2005ep})
\begin{alignat}{3} \label{csix:cases-c-1-c-2}
&\textrm{{\bf Case I: }}\qquad &C_1 & = \frac{1}{2} + \frac{i \tilde{\omega} } {2},  \quad \quad
 &C_2 & = \frac{1}{2} - \frac{i \tilde{\omega} }{2}, \nonumber \\
&\textrm{{\bf Case II: }}  &C_1 & = \frac{1}{2} + \frac{i \tilde{\omega} } {2},  \quad \quad &C_2 & =  \frac{i \tilde{\omega} }{2}, \\
&\textrm{{\bf Case III: }} &C_1 & = - \frac{i \tilde{\omega} } {2},  \quad \quad
 &C_2 & = \frac{1}{2} - \frac{i \tilde{\omega} }{2}, \nonumber \\
&\textrm{{\bf Case IV:}}  &C_1 & = - \frac{i \tilde{\omega} } {2},  \quad \quad
 &C_2 & = \frac{i \tilde{\omega} }{2}. \nonumber 
\end{alignat}

From Eqs.\ (\ref{csix:ansatz-R}), (\ref{csix:ansatz-R-y-radial}), (\ref{csix:regular solution}), and using the property of the hypergeometric functions \cite{b:DE-books}
\begin{equation} \label{csix:property-hipergeometric}
{}_{2}F_{1}(a,b;c;v) = (1-v)^{c-a-b} {}_{2}F_{1}(c-a,c-b;c;v),
\end{equation} 
we easily show that the functions $R_1$ and $R_2$ satisfy
\begin{equation} \label{csix:R_1-R_2}
|R_1|^2 - |R_2|^2 = 0
\end{equation} 
in the four cases enumerated in Eq.\ (\ref{csix:cases-c-1-c-2}), whenever $B_1=B_2=|m|$ and the oscillation frequency $\omega$ is a real number.

Therefore, from Eq.\ (37) of Ref.\ \cite{Lopez-Ortega:2004cq}, the number of fermions that cross the circle of radius $r$ per unit time is equal to
\begin{equation} \label{csix:flux for unit of time}
\frac{\partial N}{\partial t} = - 2 \pi \left( 1 -\frac{r^2}{l^2} \right)^{1/2} r \sum_m  (|R_1^{(m)}|^2 - |R_2^{(m)}|^2).
\end{equation}
This quantity vanishes because Eq.\ (\ref{csix:R_1-R_2}) holds for every $m$ when the frequency $\omega$ is a real number. From this fact follows that the radial flux of fermions from or toward the cosmological horizon is equal to zero, provided we use regular radial functions at $r=0$. When $B_1 = B_2 = |m|$, we obtain the same conclusion in the four cases enumerated in Eq.\ (\ref{csix:cases-c-1-c-2}), as can be explicitly verified.

Now, we study what happens when $B_1$ takes the value $-|m|$. For this value of $B_1$ the parameter $\tilde{c}_1$ is an integer, $\tilde{c}_1 \leq 0$; thus the first solution of the differential equation (\ref{csix:hipergeometric-v}) is \cite{b:DE-books} 
\begin{equation} \label{csix:second-option-R-1}
\hat{R}_1^{(1)} = v^{2|m|}{}_{2}F_{1}(\tilde{a}_1-\tilde{c}_1+1,\tilde{b}_1-\tilde{c}_1+1;2-\tilde{c}_1;v).
\end{equation} 

Taking $\hat{R}_1^{(1)}$ given in Eq.\ (\ref{csix:second-option-R-1}) as a part of the radial function $R_1$, we find that it is regular at $r=0$. The second solution of Eq.\ (\ref{csix:hipergeometric-v}) is \cite{b:DE-books}
\begin{align} \label{csix:hipergeometric-irregular-other-B}
\hat{R}_1^{(2)} &= v^{2 |m|}\left\{  {}_{2}F_{1} (\tilde{a}_1-\tilde{c}_1+1,\tilde{b}_1-\tilde{c}_1+1;2-\tilde{c}_1;v) \ln (v) \right. \nonumber \\
& + \frac{(1-\tilde{c}_1)! \sum_{q=1}^{1-\tilde{c}_1} \frac{(-1)^{q-1} (q-1)! \Gamma(\tilde{a}_1-\tilde{c}_1+1-q) \Gamma(\tilde{b}_1-\tilde{c}_1+1-q) }{(1-\tilde{c}_1-q)! } v^{-q}}{\Gamma(\tilde{a}_1-\tilde{c}_1+1) \Gamma(\tilde{b}_1-\tilde{c}_1+1)}  \\
& + \sum_{q=0}^{\infty} \frac{(\tilde{a}_1-\tilde{c}_1+1)_q (\tilde{b}_1-\tilde{c}_1+1)_q v^q}{q!(2-\tilde{c}_1)_q}  \left[ \psi(\tilde{a}_1-\tilde{c}_1+1+q) +  \psi(1) -\psi(\tilde{b}_1)   \right. \nonumber \\
&\left. \left. - \psi(1+q) - \psi(\tilde{a}_1) +\psi(\tilde{b}_1-\tilde{c}_1+1+q) - \psi(2-\tilde{c}_1+q) + \psi(1-\tilde{c}_1) \right]  \right\}. \nonumber
\end{align} 
Using this solution of Eq.\ (\ref{csix:hipergeometric-v}), we can verify that the radial function $R_1$ is divergent at $r=0$. Thus, when $B_1 = - |m|$, it is possible to find regular solutions at the origin of $3D$ de Sitter metric. We have an identical situation for the radial function $R_2$ when $B_2 = - |m|$.

These facts imply that there are other options for the values of $B_1$ and $B_2$, also of the case $B_1=B_2=|m|$ studied in detail in this section. The additional cases are 
\begin{alignat}{3} \label{csix:cases-B-1-B-2}
\textrm{{\bf Case II: \,\,}}\quad &B_1 & = |m|,  \quad \quad &B_2 & = -|m|, \nonumber \\
\textrm{{\bf Case III: }}\quad &B_1 & =- |m|,  \quad \quad &B_2 & = |m|,  \\
\textrm{{\bf Case IV: }}\quad &B_1 & = -|m|,  \quad \quad &B_2 & = -|m|. \nonumber 
\end{alignat}

For all cases enumerated in previous equation it is necessary to consider that the parameters $C_1$ and $C_2$ can take two different values each. Nevertheless, from the conclusions previously obtained in this section, we believe that the same result on the absorption of the fermions by the cosmological horizon of the $3D$ de Sitter spacetime will be found in the cases that we do not analyze in this paper.

\subsection{Quasinormal modes}
\label{subsection:quasinormal-modes}

The QNMs of an asymptotically flat black hole are defined as the set of oscillations of a classical field that satisfy, at the horizon and far from it, the following boundary conditions: (i) the field is purely ingoing at the horizon of the black hole; (ii) the field is purely outgoing at infinity \cite{Kokkotas:1999bd}. If the black hole is not asymptotically flat, then it is possible to impose other boundary conditions far from the horizon; see for example Refs.\ \cite{Birmingham:2001pj}, \cite{Lopez-Ortega:2005ep}, \cite{Fernando:2003ai}, \cite{Cardoso:2001hn}. For static and stationary black holes the QNMs for different fields have been studied mainly using numerical methods \cite{Kokkotas:1999bd}.

An interesting problem is to investigate the analogous of the QNMs of black holes for the fields that are propagating in de Sitter metric. Clearly, the definition of the de Sitter QNMs must be different from that used in the case of a black hole. In this paper, following Ref.\ \cite{Choudhury:2003wd} (see also Refs.\ \cite{Brady:1999wd}, \cite{Natario:2004jd}), the de Sitter QNMs boundary conditions are: (i) the field is regular at the origin; (ii) the field is purely outgoing near the cosmological horizon.

Using the exact solutions of the Dirac equation previously found, we can calculate the QNM frequencies of a massless Dirac field propagating in $3D$ de Sitter background. In the following we study in detail the case $B_1=B_2=|m|$, $C_1=\tfrac{1}{2}+\tfrac{i \tilde{\omega}}{2}$, and $C_2=\tfrac{i \tilde{\omega}}{2}$.

By the property of the hypergeometric functions \cite{b:DE-books}
\begin{eqnarray} \label{csix:hipergeometric-property-z-1-z}
{}_2F_1(a,b;c;v) = \frac{\Gamma(c) \Gamma(c-a-b)}{\Gamma(c-a) \Gamma(c - b)} {}_2 F_1 (a,b;a+b+1-c;1-v) \hspace{1cm} \nonumber \\
+ \frac{\Gamma(c) \Gamma( a +b - c)}{\Gamma(a) \Gamma(b)} (1-v)^{c-a -b} {}_2F_1(c-a, c-b; c + 1 -a-b; 1 -v),
\end{eqnarray}
which is valid when $c-a-b$ is not an integer, we can write the radial functions $R_1$ and $R_2$ that are regular at the origin in the form
\begin{align} \label{ds3:radial-1}
R_1 &= \frac{2-v}{\sqrt{2}}v^{|m|-\tfrac{1}{2}} \left\{(1-v)^{C_1-\tfrac{1}{4}} \frac{\Gamma(\tilde{c_1}) \Gamma(\tilde{c_1}-\tilde{a_1}- \tilde{b_1})}{\Gamma(\tilde{c_1}-\tilde{a_1}) \Gamma(\tilde{c_1} - \tilde{b_1})}  \right. \\ 
& \times \left. {}_2F_1(\tilde{a_1},\tilde{b_1};1+\tilde{a_1}+\tilde{b_1}-\tilde{c_1};1-v)  + (1-v)^{\tfrac{1}{4}-C_1}  \frac{\Gamma(\tilde{c_1}) \Gamma(\tilde{a_1} + \tilde{b_1} -\tilde{c_1})}{\Gamma(\tilde{a_1}) \Gamma(\tilde{b_1})}  \nonumber \right. \\
&\left. \times {}_2F_1(\tilde{c_1}-\tilde{a_1},\tilde{c_1} - \tilde{b_1};1 - \tilde{a_1} - \tilde{b_1} + \tilde{c_1};1-v)  \right\},\nonumber
\end{align}
and a similar expression for $R_2$ which can be easily found from Eq.\ (\ref{ds3:radial-1}) substituting the subscript ``$1$'' of the parameters  by ``$2$''.

According to the definitions previously given, the following relation holds
\begin{equation} \label{csix:v-tortoise-relation}
1-v={\rm exp}(-\tfrac{2 r_*}{l}),
\end{equation} 
where 
\begin{equation} \label{eq:sec2:tortoise}
r_* = \frac{l}{2} \ln \frac{1+z}{1-z}
\end{equation}
stands for the usual tortoise coordinate which satisfies $r_* \to 0$ as $r \to 0$ and $r_* \to \infty $ as $r \to l$ \cite{Lopez-Ortega:2004cq}.

Employing Eqs.\ (\ref{separable form}) and (\ref{csix:v-tortoise-relation}), we get that the second term in curly braces in the right-hand side of Eq.\ (\ref{ds3:radial-1}) represents an outgoing wave, whereas the first term in curly braces represents an ingoing wave. Thus, in order to satisfy the definition of the QNMs in de Sitter spacetime, we must cancel the first term in curly braces of Eq.\ (\ref{ds3:radial-1}). One way to achieve this cancellation is to find the poles of the terms $\Gamma(\tilde{c_{1}} - \tilde{a_{1}})$ and $\Gamma(\tilde{c_{1}} - \tilde{b_{1}})$ \cite{Choudhury:2003wd}, \cite{Brady:1999wd}. These poles are localized at \cite{b:DE-books} 
\begin{equation} \label{dS3-conditions-quasi}
\tilde{c_{1}} - \tilde{a_{1}} = -n_{1}, \qquad \textrm{and} \qquad \tilde{c_{1}} - \tilde{b_{1}} = -n_{1},
\end{equation} 
respectively, where $n_{1}=0,1,2,\dots $

From the values of $\tilde{a_{1}}$, $\tilde{b_{1}}$, and $\tilde{c_{1}}$ given in Eq.\ (\ref{csix:abc-hipergeometric-v}), the previous equations yield 
\begin{equation} \label{dS3:condition-quasinormal}
|m| = -n_{1}, \qquad \quad \qquad  |m| + \frac{1}{2} - i \tilde{\omega} = -n_{1},
\end{equation} 
respectively. We cannot satisfy the first relation in Eqs.\ (\ref{dS3:condition-quasinormal}) (that is, $\Gamma(\tilde{c_1} - \tilde{b_1})$ has no poles), but the second relation in (\ref{dS3:condition-quasinormal}) implies
\begin{equation} \label{dS3-values-frequencies}
i \tilde{\omega} = |m| + \frac{1}{2} + n_{1} .
\end{equation} 

Note that the term $\Gamma(\tilde{c_1})$ does not have poles, whereas the poles of the term $\Gamma(\tilde{c_1}-\tilde{a_1}- \tilde{b_1})$ are localized at
\begin{equation}
\tilde{c_1}-\tilde{a_1}- \tilde{b_1} = -n_{1}.
\end{equation}
Using the values of $\tilde{a}_1$, $\tilde{b}_1$, and $\tilde{c}_1$ we find that the previous equation becomes
\begin{equation} \label{dS3:other-pole}
i \tilde{\omega} = n_{1} - \frac{1}{2}.
\end{equation} 

Since the quantity $m$ is a half-integer (see Appendix \ref{appendix:half-integer-m}), the expression in the right-hand side of Eq.\ (\ref{dS3-values-frequencies}) is an integer, whereas the corresponding term in Eq.\ (\ref{dS3:other-pole}) is a half-integer; therefore they cannot be equal, and there is no pole cancellation in the first factor of the expansion of the radial function (\ref{ds3:radial-1}), (see Refs.\ \cite{Choudhury:2003wd}, \cite{Brady:1999wd}, \cite{Natario:2004jd}). We notice that for the frequencies $i \tilde{\omega}$ given in Eq.\ (\ref{dS3-values-frequencies}) the quantity $\tilde{c_1}-\tilde{a_1}- \tilde{b_1}$ is not an integer, that is, if the quantity $m$ is a half-integer, then there is no pole cancellation.

By analogy with the previous case, the radial function $R_2$ will be purely outgoing as $r \to l$ if either of the conditions
\begin{equation}\label{dS3-conditions-quasi-2}
\tilde{c_{2}} - \tilde{a_{2}} = -n_{1}, \qquad \quad \textrm{or} \qquad \quad \tilde{c_{2}} - \tilde{b_{2}} = -n_{1}, 
\end{equation} 
holds. In previous equations we can satisfy the first condition but not the second one. The QNM frequencies obtained are equal to those given in Eq.\ (\ref{dS3-values-frequencies}). If $B_1 = B_2 = |m|$, then for the other possible values of $C_1$ and $C_2$ (see Eq.\ (\ref{csix:cases-c-1-c-2})), we also find the QNM frequencies (\ref{dS3-values-frequencies}).

Taking into account that the exact solutions of the massless Dirac equation studied in the present section have a harmonic time dependence of the form $\textrm{e}^{-i \omega t}$, we find that for the QNM frequencies (\ref{dS3-values-frequencies}) the amplitude of the field diminishes with time.

When a massless Dirac field has some of the frequencies (\ref{dS3-values-frequencies}), an explicit calculation of the radial flux (\ref{csix:flux for unit of time}) shows that it is different from zero. Therefore the modes with frequencies (\ref{dS3-values-frequencies}) are QNMs.

The frequencies (\ref{dS3-values-frequencies}) are equal to those given in Eqs.\ (39) and (40) of Ref.\ \cite{Du:2004jt} when the mass of the Dirac field is equal to zero and if the quantity $\ell$, which appears in Eqs.\ (39) and (40) of Ref.\ \cite{Du:2004jt}, is a half-integer greater than zero (Eq.\ (39)) or smaller than zero (Eq.\ (40)). The exact values allowed for $\ell$ are not explicitly found in Ref.\ \cite{Du:2004jt}. 

The existence of the QNM frequencies for a massless Dirac field implies that for this case the conjecture of Ref.\ \cite{Natario:2004jd} holds. For a massless classical field, this conjecture proposes that there are well defined de Sitter QNMs only in odd spacetime dimensions. (See below.)

As previously discussed in Ref.\ \cite{Natario:2004jd}, the boundary conditions imposed in Ref.\ \cite{Myung:2003ki} on a classical field propagating in de Sitter spacetime to calculate their QNM frequencies are not equal to those imposed in Refs.\ \cite{Choudhury:2003wd}, \cite{Brady:1999wd}, \cite{Natario:2004jd}, \cite{Abdalla:2002hg} (and in the present paper). In \cite{Myung:2003ki}, the de Sitter QNMs boundary conditions are: (i) the field has zero flux at $r=0$; (ii) the field has a outgoing flux different from zero near $r=l$. Moreover, in Ref.\ \cite{Myung:2003ki} only real frequencies are studied, therefore it is deduced that the de Sitter QNM frequencies do not exist. As is well known, the QNM frequencies are complex numbers. For these reasons we do not compare our results with those of Ref.\ \cite{Myung:2003ki}.

\section{Massless fields in $4D$ de Sitter spacetime}
\label{csix:dS4-massless-fields}

The behavior of the classical fields in $4D$ curved spacetimes can be analyzed with an efficient method that employs the Newman-Penrose formalism \cite{RS:Newman-1962-BWM-RS}, \cite{Chandrasekhar book}, \cite{Teukolsky:1973ha} in order to get the equations of motion for the fields in a suitable form. In this section we study some aspects of the dynamics of massless fields propagating in de Sitter metric in four dimensions using this method.

Henceforth we use the null tetrad\footnote{In this work $z^*$ stands for the complex conjugate of $z$.} \cite{RS:Newman-1962-BWM-RS}, \cite{Chandrasekhar book}
\begin{align} \label{csix:tetrad NP}
l^\nu &= \frac{1}{\sqrt{2}}\left(\frac{1}{P},P,0,0 \right), \nonumber \\
n^\nu &= \frac{1}{\sqrt{2}}\left(\frac{1}{P},-P,0,0 \right), \\
m^\nu &= \frac{1}{\sqrt{2}r}\left( 0,0,1, i \csc \theta \right), \nonumber \\
m^{\nu *} &= \frac{1}{\sqrt{2}r}\left( 0,0,1, - i \csc \theta \right), \nonumber
\end{align} 
where $P^2$ was previously defined in Eq.\ (\ref{csix:definition P}). The tetrad (\ref{csix:tetrad NP}) is different from that used in Refs.\ \cite{Suzuki:1995nh}, \cite{Otchik:1985ih}; however, we employ (\ref{csix:tetrad NP}) because the necessary algebraic steps are simpler than in the tetrad used in  \cite{Suzuki:1995nh}, \cite{Otchik:1985ih}. 

The spin coefficients for the tetrad (\ref{csix:tetrad NP}) are equal to\footnote{For the spin coefficients $\mu$ and $\pi$ in the original Newman-Penrose notation \cite{RS:Newman-1962-BWM-RS}, in the present paper we use the symbols $\tilde{\mu}$ and $\tilde{\pi}$ respectively.}
\begin{align} \label{csix:espin-coeff}
\kappa = \sigma &= \lambda = \nu = \tau = \tilde{\pi} = 0, \nonumber \\
&\rho = \tilde{\mu} = - \frac{P}{\sqrt{2} r}, \nonumber \\
&\epsilon = \gamma = -\frac{r}{2 \sqrt{2} l^2 P}, \\
&\alpha = - \beta = -\frac{\cot \theta}{2 \sqrt{2} r}. \nonumber
\end{align} 
For the Weyl tensor, the traceless part of the Ricci tensor, and the scalar curvature we get
\begin{align} \label{csix:Weyl-Ricci-curvature}
\Psi_0=\Psi_1=&\Psi_2=\Psi_3=\Psi_4=0, \nonumber \\
\Phi_{00}=\Phi_{01}=\Phi_{02}&=\Phi_{11}=\Phi_{12}=\Phi_{22} = 0 ,\\
&\Lambda = \frac{1}{2 l^2}.\nonumber 
\end{align}

As is well known, in a type D or simpler spacetime (in particular de Sitter background), the equations of motion for the massless fields of spin $s =  \tfrac{1}{2}, 1, 2$ admit separable solutions \cite{Chandrasekhar book}, \cite{Teukolsky:1973ha}, \cite{PDK:Dudley-eq-sep-PD}. The equations for the field components invariant under gauge transformations and infinitesimal rotations of the tetrad are given by \cite{Chandrasekhar book}, \cite{Teukolsky:1973ha}
\begin{eqnarray} \label{csix:chi+-motion-equation}
\{[D -(2s -1)\epsilon + \epsilon^* -2s\rho - \rho^*][\Delta + \tilde{\mu} - 2s\gamma] - (2s-1)(s-1)\Psi_2 \nonumber \\
-[\delta + \tilde{\pi}^* - 2 s \tau - \alpha^* - (2s -1)\beta][\delta^* + \tilde{\pi} - 2s \alpha] \} \chi_s^{+}=0,
\end{eqnarray} 
for $s=+\tfrac{1}{2}$, $+1$, $+2$, and
\begin{eqnarray} \label{csix:chi-motion-equation}
\{ [\Delta -(2s+1)\gamma -\gamma^* - 2 s \tilde{\mu} + \tilde{\mu}^*][D -2s\epsilon - \rho] - (2s+1)(s+1)\Psi_2 \nonumber \\ 
-[\delta^* -\tau^* -2s\tilde{\pi} + \beta^* -(2s+1)\alpha][\delta - \tau -2s\beta ]\}\chi_s^{-}=0,
\end{eqnarray} 
for $s=-\tfrac{1}{2}$, $-1$, $-2$.

In $4D$ de Sitter metric, using the tetrad (\ref{csix:tetrad NP}) and proposing the ansatz 
\begin{equation} \label{csix:ansatz-chi+}
\chi_s^+ = {\rm e}^{-i\omega t} R_s^+(r) \,{}_s Y_{jm}(\theta, \phi),
\end{equation} 
where the symbols ${}_s Y_{jm}$ stand for the spin-weighted spherical harmonics \cite{csix:spin-harmonic-harmonic}, we can show that Eq.\ (\ref{csix:chi+-motion-equation}) simplifies to 
\begin{eqnarray} \label{csix:differential-equation-z-R+}
\left[\frac{{\rm d}^2}{{\rm d}z^2} - \left(\frac{2z}{1-z^2} - \frac{2(s+1)}{z} \right) \frac{{\rm d}}{{\rm d}z} + \frac{\tilde{\omega}^2}{(1-z^2)^2} + \frac{2 i \tilde{\omega} s}{z(1-z^2)^2} + \frac{2s}{z^2} \right. \nonumber \\
\left.-\frac{2s^2+s+2}{1-z^2} -\frac{s^2z^2}{(1-z^2)^2}-\frac{(j+s)(j-s+1)}{z^2(1-z^2)} \right]R_s^+ = 0,
\end{eqnarray} 
where $\tilde{\omega}$ and $z$ are defined as in the previous section.

Employing the variable $y$ defined in Eq.\ (\ref{csix:y-definition}), the differential equation (\ref{csix:differential-equation-z-R+}) becomes
\begin{eqnarray}
\left\{\frac{{\rm d}^2}{{\rm d}y^2} +\left[\frac{1}{y}-\frac{2(s+1)}{1-y} -\frac{2(s+1)}{1+y}\right]\frac{{\rm d}}{{\rm d}y} + \frac{(\tilde{\omega}+i s)^2}{4y^2}  \hspace{1.4cm} \right. \nonumber \\
+ \frac{i \tilde{\omega}s}{y(1-y)} -\frac{j(j+1)-s(s+1)}{y(1-y)}-\frac{j(j+1)-s(s+1)}{(1-y)^2} \hspace{.4cm} \nonumber \\ 
\left. \qquad \hspace{.1cm}-\frac{(s+2)(s+1)}{y (1+y)} +\frac{(s+2)(s+1)}{(1+y)^2}  \right\}R_s^+ = 0 .
\end{eqnarray} 
Proposing that the function $R_s^+$ takes the form\footnote{In Ref.\ \cite{Suzuki:1995nh} was shown that the radial differential equations of the massless fields of spin  $s=0$, $\tfrac{1}{2}, 1, 2$ moving in de Sitter background can be simplified to differential equations of hypergeometric type. The tetrad used in Ref.\ \cite{Suzuki:1995nh} is different from that used in the present paper, and we believe that the method used here to make the algebraic simplification is even simpler, (see also Section \ref{csix:dS4-massive-Dirac}).}
\begin{equation} \label{csix:R+-radial}
R_s^+ = (1+y)^{A^+}(1-y)^{B^+}y^{C^+} \hat{R}^+_s ,
\end{equation} 
with the parameters
\begin{align} \label{csix:constant-A+B+C+}
A^+ &= 1 + |s|, \nonumber \\
B^+ &= \left\{ \begin{array}{l} j - |s|, \\ - (j+|s|+1), \end{array} \right. \\
C^+ & = \left\{  \begin{array}{l} \frac{i\tilde{\omega}}{2} -\frac{|s|}{2}, \\ -\frac{i\tilde{\omega}}{2} + \frac{|s|}{2}, \end{array} \right. \nonumber 
\end{align} 
we find that the function $\hat{R}^+_s$ must be a solution of
\begin{equation} \label{csix:4D-hipergeometric+}
\left[v(1-v) \frac{{\rm d}^2}{{\rm d}v^2} + (c^+ - (a^+ + b^+ + 1)v)\frac{{\rm d}}{{\rm d}v} -a^+ b^+  \right]\hat{R}_s^+ = 0,
\end{equation} 
with $v=1-y$, as in the previous section, and
\begin{align} \label{csix:a-b-c-massless-fields-4D}
a^+ &= C^+ + B^+  + 1 + \frac{i\tilde{\omega} }{2} + \frac{3|s|}{2},\nonumber \\
b^+ &= C^+ + B^+  + 1 - \frac{i\tilde{\omega} }{2} + \frac{|s|}{2}, \\
c^+ &= 2 (B^+ +  |s| + 1).\nonumber
\end{align} 

Taking into account the possible values for $B^+$ given in Eq.\ (\ref{csix:constant-A+B+C+}), we see that $c^+$ is a positive (for $B^+ = j - |s|$) or negative integer (for $B^+ = - (j + |s| + 1)$). As in the previous section, in both cases, it is possible to find solutions of Eq.\ (\ref{csix:4D-hipergeometric+}) that lead to regular radial functions $R_s^+$ at the origin of de Sitter spacetime. Moreover, in both cases there are solutions of Eq.\ (\ref{csix:4D-hipergeometric+}) that lead to radial functions $R_s^+$ which are divergent at $r=0$.

In the following we only study in detail the case $B^+ = j - |s|$. For the other case (when $B^+ = - (j + |s|+1)$) we expect to find identical physical results. If $B^+ = j - |s|$, then the quantity $c^+$ is a positive integer. Therefore the solutions of the differential equation (\ref{csix:4D-hipergeometric+}) are \cite{b:DE-books} 
\begin{align} \label{csix:4D-regular-solution}
\hat{R}_s^{+(1)} &= {}_{2}F_{1}(a^+,b^+;c^+;v), \qquad \qquad \\
\hat{R}_s^{+(2)} &= {}_{2}F_{1}(a^+,b^+;c^+;v) \ln (v) \nonumber \\
&+ \frac{(c^+ - 1)!}{\Gamma(a^+) \Gamma(b^+)} \sum_{q=1}^{c^+ - 1} (-1)^{q-1} (q - 1)! \frac{\Gamma(a^+ - q) \Gamma(b^+ -q)}{(c^+ -q - 1)!} v^{-q} \nonumber \\
& + \sum_{q=0}^{\infty} \frac{(a^+)_q (b^+)_q}{q! (c^+)_q} v^q \left[\psi(a^+ + q) + \psi(b^+ + q) - \psi(c^+ + q) - \psi(1 + q) \right. \nonumber \\
 & \left. \hspace{0cm} - \psi(a^+ -  c^+ +1) - \psi(b^+ - c^+ + 1) + \psi(1) + \psi(c^+ -1) \right].
\end{align} 

The second solution $\hat{R}_s^{+(2)}$ of Eq.\ (\ref{csix:4D-hipergeometric+}) leads to a radial function $R^+_s$ which is divergent at the origin. Therefore in the rest of the present section we only consider the first solution $\hat{R}_s^{+(1)}$ of the differential equation (\ref{csix:4D-hipergeometric+}), which hereafter we simply denote as $\hat{R}_s^+$.

A similar procedure with 
\begin{align} \label{csix:R--radial-several}
\chi_s^- &= {\rm e}^{-i\omega t} R_s^-(r)\, {}_s Y_{jm}(\theta, \phi),\nonumber \\
R_s^- &= (1+y)^{A^-}(1-y)^{B^-}y^{C^-} \hat{R}^-_s, \\
A^- &= 1 + |s|,  \nonumber \\
B^- &= \left\{ \begin{array}{l} j - |s|, \\ - (j+|s|+1), \end{array} \right. \nonumber \\
C^- & = \left\{  \begin{array}{l} \frac{i\tilde{\omega}}{2} + \frac{|s|}{2}, \\ -\frac{i\tilde{\omega}}{2} - \frac{|s|}{2}, \end{array} \right. \nonumber 
\end{align}
simplifies Eq.\ (\ref{csix:chi-motion-equation}) to a differential equation of hypergeometric type for the function $\hat{R}_s^-$,
\begin{equation} \label{csix:4D-hipergeometric-}
\left[v(1-v) \frac{{\rm d}^2}{{\rm d}v^2} + (c^- - (a^- + b^- + 1)v)\frac{{\rm d}}{{\rm d}v} -a^- b^-  \right]\hat{R}_s^- = 0,
\end{equation} 
where
\begin{align}
a^- &= C^- + B^-  + 1 + \frac{i\tilde{\omega} }{2} + \frac{|s|}{2},\nonumber \\
b^- &= C^- + B^-  + 1 - \frac{i\tilde{\omega} }{2} + \frac{3|s|}{2}, \\
c^- &= 2 (B^- +  |s| + 1).\nonumber
\end{align}

Depending on the value chosen for $B^-$, the parameter $c^-$ is a positive or negative integer. In both cases there are solutions which are regular at the origin of de Sitter metric. In the rest of the present section, we only study in detail the case $B^- = j - |s|$, that is, when the quantity $c^-$ is a positive integer.

If $B^-=j - |s|$, then making a similar analysis for $R^-_s$, we find that the solution of the differential equation (\ref{csix:4D-hipergeometric-}) 
\begin{equation}
\hat{R}_s^- = {}_{2}F_{1}(a^-,b^-;c^-;v)
\end{equation} 
leads to a radial function $R^-_s$ which is regular at $r=0$.

As in Section \ref{csix:dS3-Dirac-massless}, for  given $B^+$ and $B^-$, it is necessary to note that the parameters $C^+$ and $C^-$ can take two values each. This fact implies that we have the cases (see also Refs.\ \cite{Abdalla:2002hg}, \cite{Lopez-Ortega:2005ep})
\begin{alignat}{3} \label{csix:4D-cases-C-1-C-2}
&\textrm{{\bf Case I: }}\qquad &C^+ & = - \frac{|s|}{2} + \frac{i \tilde{\omega} } {2},  \quad \quad
 &C^- & = + \frac{|s|}{2} + \frac{i \tilde{\omega} }{2}, \nonumber \\
&\textrm{{\bf Case II: }}  &C^+ & = - \frac{|s|}{2} + \frac{i \tilde{\omega} } {2},  \quad \quad &C^- & =  - \frac{|s|}{2} - \frac{i \tilde{\omega} }{2}, \\
&\textrm{{\bf Case III: }} &C^+ & = +\frac{|s|}{2} - \frac{i \tilde{\omega} } {2},  \quad \quad
 &C^- & = +\frac{|s|}{2} + \frac{i \tilde{\omega} }{2}, \nonumber \\
&\textrm{{\bf Case IV:}}  &C^+ & =+ \frac{|s|}{2} - \frac{i \tilde{\omega} } {2},  \quad \quad
 &C^- & = - \frac{|s|}{2} - \frac{i \tilde{\omega} }{2}. \nonumber 
\end{alignat}

If the frequency $\omega$ is a real number and the radial functions are regular at the origin of de Sitter spacetime, when  $B^+ = B^- = j - |s|$ it is possible to show that the relation 
\begin{equation} \label{csix:null-flux}
|R_s^+|^2 - |R_s^-|^2 = 0
\end{equation}
holds in the four cases enumerated in Eq.\ (\ref{csix:4D-cases-C-1-C-2}) for $s=\tfrac{1}{2}$, $1$, $2$.

\subsection{Flux calculation}
\label{csix:subsection-flux}

In this subsection we calculate the radial flux of the massless fields of spin $s=\tfrac{1}{2}$, $1$, with regular radial functions at the origin of de Sitter metric. The analysis for a scalar field ($s=0$) was published in Refs.\ \cite{Myung:2003cn}, \cite{Myung:2003ki}.

If the spin of the massless field is $s=\tfrac{1}{2}$, then we use that the radial component of the conserved current for a massless Dirac field in the tetrad (\ref{csix:tetrad NP}) is equal to \cite{Chandrasekhar book}
\begin{equation} \label{csix:Radial-component-J}
J^r = P \left(|\chi_{\tfrac{1}{2}}^-|^2 - |\chi_{\tfrac{1}{2}}^+|^2\right).
\end{equation} 
Therefore the number of fermions that cross the sphere of radius $r$ is given by \cite{Chandrasekhar book}
\begin{equation} \label{csix:Weyl-flux-equal-zero}
\frac{\partial N}{\partial t} = - \int_0^{2\pi} \int_0^{\pi} J^r \sqrt{|g|}\,\, {\rm d} \theta \,\,{\rm d} \phi \propto   P r^2 \left(|R_{\tfrac{1}{2}}^+|^2 - |R_{\tfrac{1}{2}}^-|^2\right).
\end{equation} 
This quantity is equal to zero due to Eq.\ (\ref{csix:null-flux}).

Using Eq.\ (339) given in Chapter 5 of Ref.\ \cite{Chandrasekhar book}, we find that the component $T^r_{\,\,t}$ of the energy-momentum tensor of an electromagnetic field propagating in $4D$ de Sitter background is equal to
\begin{equation} \label{csix:T-r-t-electromagnetic}
T^r_{\,\,t} = \frac{P}{4 \pi}\left(|\chi_1^-|^2 - |\chi_1^+|^2\right)
\end{equation} 
when we use the tetrad (\ref{csix:tetrad NP}).

Integrating Eq.\ (\ref{csix:T-r-t-electromagnetic}) over the coordinates $\theta$ and $\phi$, we find that the energy flux of an electromagnetic field takes the form
\begin{equation} \label{csix:em-flux-equal-zero}
\frac{{\rm d}E}{{\rm d}t} \propto r^2 P \left( |R_1^+|^2 - |R_1^-|^2 \right),
\end{equation} 
which is equal to zero owing to Eq.\ (\ref{csix:null-flux}). Thus for a massless Dirac field and an electromagnetic field moving in de Sitter spacetime Eqs.\ (\ref{csix:Weyl-flux-equal-zero}) and (\ref{csix:em-flux-equal-zero}) show that their radial flux is equal to zero whenever the radial functions are regular at $r=0$ and the frequency $\omega$ is a real number. This conclusion holds in the four cases enumerated in Eq.\ (\ref{csix:4D-cases-C-1-C-2}). 

However, we have not been able to prove the equivalent result for the gravitational perturbations. By using the Isaacson  stress-energy tensor \cite{dS3D-Isaacson}, we found a complicated expression for the energy momentum flux at any point in de Sitter metric. Although we can use Eqs.\ (\ref{csix:chi+-motion-equation}) and (\ref{csix:chi-motion-equation}) to describe the dynamics of a massless field of spin $s=\tfrac{3}{2}$ (Rarita-Schwinger) in $4D$ de Sitter spacetime \cite{Aichelburg:1980up}, we do not know an expression for its energy-momentum tensor that allows us to calculate its radial flux.

\subsection{Quasinormal modes}
\label{csix:subsection:quasinormal-modes}

By the definition of de Sitter QNMs previously given in Subsection \ref{subsection:quasinormal-modes}, it is clear that the solutions of the equations of motion for the massless fields found in the present section allow us to calculate these by means of a similar procedure to that of Refs.\ \cite{Choudhury:2003wd}, \cite{Brady:1999wd}, \cite{Natario:2004jd}. In the following we study in detail the case $B^+ = j - |s|$ and $C^+=\tfrac{i \tilde{\omega}}{2} - \tfrac{|s|}{2}$. For these values of the parameters $B^+$ and $C^+$, the regular solution at $r=0$ is 
\begin{equation} \label{csix:radial-plus-complete}
R^+_s = (2-v)^{1+|s|} v^{j-|s|}(1-v)^{\tfrac{i \tilde{\omega}}{2} - \tfrac{|s|}{2}} {}_2 F_1(a^+,b^+;c^+;v).
\end{equation} 

Employing the property of the hypergeometric functions (\ref{csix:hipergeometric-property-z-1-z}) valid when $c-a-b$ is not an integer \cite{b:DE-books}, we can write the function $R^+_s$ of the previous equation in the form
\begin{align} \label{csix:R-+-relation-v-1-v}
& R^+_s = (2-v)^{1+|s|} v^{j-|s|}  \left\{ (1-v)^{\tfrac{i \tilde{\omega}}{2} - \tfrac{|s|}{2}}  \frac{\Gamma(c^+) \Gamma(c^+ -a^+ -b^+)}{\Gamma(c^+ -a^+) \Gamma(c^+ - b^+)} \right. \nonumber \\ 
&\times {}_2 F_1 (a^+,b^+;a^+ + b^+ +1-c^+;1-v) \,\, + \,\, (1-v)^{-\tfrac{i \tilde{\omega}}{2} + \tfrac{|s|}{2}}   \\ 
&\left. \times \frac{\Gamma(c^+) \Gamma( a^+ +b^+ - c^+)}{\Gamma(a^+) \Gamma(b^+)} {}_2F_1(c^+ -a^+, c^+ -b^+; c^+ + 1 -a^+ -b^+; 1 -v)  \right\}. \nonumber
\end{align} 

From Eq.\ (\ref{csix:ansatz-chi+}), we note that the massless fields studied here also have an harmonic dependence on time $\textrm{e}^{-i \omega t}$; therefore the first term in curly braces of Eq.\ (\ref{csix:R-+-relation-v-1-v}) represents an ingoing wave as $r \to l$, whereas the second term represents an outgoing wave as $r \to l$. To satisfy the boundary conditions of the de Sitter QNMs, we look for the poles of the terms $\Gamma(c^+ -a^+)$ and $\Gamma(c^+ - b^+)$ (see Subsection \ref{subsection:quasinormal-modes}). These poles are localized at \cite{b:DE-books}
\begin{equation} \label{csix:conditions-quasinormal}
c^+ - a^+ = -n_1, \qquad \quad c^+ - b^+ = -n_1, \qquad n_1=0,1,2,\dots 
\end{equation} 
 
Hence the quantities $c^+ - a^+$ and $c^+ - b^+$ must be negative integers, but from the values of $a^+$, $b^+$, and $c^+$ given in Eqs.\ (\ref{csix:a-b-c-massless-fields-4D}), we see that if $c^+-a^+-b^+ = -(i\tilde{\omega}-|s|)$ is not an integer, then the quantities $c^+ - a^+=j+1-|s|-(i \tilde{\omega} -|s|)$ and $c^+ - b^+=j+1+|s|$ cannot be negative integers. That is, we cannot satisfy the conditions (\ref{csix:conditions-quasinormal}). Therefore, if the quantity $c^+-a^+-b^+$ is not an integer, then there are no QNM frequencies for the massless fields of spin $s=\tfrac{1}{2},1,2$.

In Refs.\ \cite{Choudhury:2003wd}, \cite{Brady:1999wd} this fact was not noticed; moreover, in these papers there is the so called pole cancellation, which implies that a massless minimally coupled scalar field has no well defined QNM frequencies when it is propagating in $4D$ de Sitter spacetime. Something similar happens in Ref.\ \cite{Natario:2004jd} for the gravitational perturbations if the dimension $D$ of de Sitter spacetime is even, $D \geq 4$. (See Appendix \ref{appendix:quasinormal-modes-4D}.)

We stress that if the factor $\Gamma(c^+-a^+-b^+)$ has poles, then Eq.\ (\ref{csix:hipergeometric-property-z-1-z}) is not valid \cite{b:DE-books}. When $c^+-a^+-b^+= -n$, $n=1,2,\dots$, instead of Eq.\ (\ref{csix:hipergeometric-property-z-1-z}), we must use\footnote{If $c^+-a^+-b^+=n$ or $c^+-a^+-b^+=0$, $n=1,2,3,\dots$, then there are no QNM frequencies because the amplitude of the purely outgoing fields increases with time or the fields have no time dependence.\label{footnote-9}} \cite{b:DE-books}
\begin{align} \label{eq:hypergeometric-c-a-b-integer}
{}_2F_1 (a,b;c;y) &=   \frac{\Gamma(a+b-n) \Gamma(n)}{\Gamma(a)\Gamma(b)} (1-y)^{-n} \sum_{q=0}^{n-1}\frac{(a-n)_q (b-n)_q}{q! (1-n)_q}(1-y)^q  \nonumber \\ 
& - \frac{(-1)^n \Gamma(a+b-n)}{\Gamma(a-n)\Gamma(b-n)} \sum_{q=0}^\infty \frac{(a)_q (b)_q}{q!(n+q)!}(1-y)^q [\textrm{In}(1-y)  \nonumber \\
&  -\psi(q+1) -\psi(q+n+1)+\psi(a+q)+\psi(b+q)] .
\end{align} 

From Eq.\ (\ref{eq:hypergeometric-c-a-b-integer}), we find that the function $R^+_s$ takes the form
\begin{align}
R^+_s &= (2-v)^{1+|s|} v^{j-|s|} (1-v)^{\tfrac{n}{2}} \nonumber \\
&\times \left\{ \frac{\Gamma(a^+ +b^+ -n) \Gamma(n)}{\Gamma(a^+)\Gamma(b^+)} (1-v)^{-n} \sum_{q=0}^{n-1}\frac{(a^+-n)_q (b^+-n)_q}{q! (1-n)_q}(1-v)^q  \right. \nonumber \\ 
& - \frac{(-1)^n \Gamma(a^+ +b^+ -n)}{\Gamma(a^+ -n)\Gamma(b^+ -n)} \sum_{q=0}^\infty \frac{(a^+)_q (b^+)_q}{q!(n+q)!}(1-v)^q [\textrm{In}(1-v)-\psi(q+1)  \nonumber \\
& \left.  - \frac{{}}{{}} \psi(q+n+1)+\psi(a^+ +q)+\psi(b^+ +q)] \right\}.
\end{align}
In this expression for $R^+_s$, the second term in curly braces represents an ingoing wave as $r\to l$; then, in order to have a purely outgoing field as $r \to l$, it is necessary to satisfy the condition
\begin{equation} \label{eq:conditions-quasinormal-integer}
a^+ - n = -n_1, \qquad \textrm{or} \qquad b^+ - n = -n_1.
\end{equation} 
From the values for $a^+,b^+$, and $c^+$ given in Eq.\ (\ref{csix:a-b-c-massless-fields-4D}) we cannot satisfy the first condition in Eq.\ (\ref{eq:conditions-quasinormal-integer}), but from the second condition we find that the QNM frequencies are equal to
\begin{equation} \label{eq-quasinormal-frequencies-4D}
i \tilde{\omega} = j+1+n_1.
\end{equation} 

Making a similar analysis for $R^-$, when $B^- = j - |s| $ and $C^- = \tfrac{i\tilde{\omega}}{2} + \tfrac{|s|}{2}$, we also find the  de Sitter QNM frequencies (\ref{eq-quasinormal-frequencies-4D}). We point out that for the frequencies (\ref{eq-quasinormal-frequencies-4D}) the amplitude of the field decreases with time. For massless fields of spin $s=\tfrac{1}{2}, 1$ with frequencies given by Eq.\ (\ref{eq-quasinormal-frequencies-4D}), an explicit calculation of the radial flux shows that it is different from zero for each mode.

The dynamics of the gravitational perturbations propagating in $D$-dimensional de Sitter metric ($D\geq4$) has been studied in Ref.\ \cite{Natario:2004jd} using a different set of exact solutions. Moreover, the authors of Ref.\ \cite{Natario:2004jd} conjectured that the massless classical fields moving in de Sitter background do not have QNM frequencies when the spacetime dimension is even. The results presented here show that this conjecture is not valid in four dimensions for the massless fields of spin $\tfrac{1}{2}$, $1$, $2$. In particular our result for the QNM frequencies of the gravitational perturbations propagating in $4D$ de Sitter metric is different from that of \cite{Natario:2004jd}.

To finish the present section we note that it is possible to consider other permutations for the values of $B^+$ and $B^-$, different from the combination studied in detail here; for example, the combination $B^+=j-|s|$ and $B^-=-(j+|s|+1)$  (and other two options). Nevertheless, we believe that in the other cases the same results hold for the value of the cross section of the cosmological horizon and for the values of the de Sitter QNM frequencies.

\section{Massive Dirac field in $4D$ de Sitter spacetime}
\label{csix:dS4-massive-Dirac}

When the background spacetime is $4D$ de Sitter in static coordinates, some time ago Otchik found exact solutions of the differential equations system for the radial functions of the massive Dirac equation \cite{Otchik:1985ih} (see Appendix \ref{appendix:solution-dirac-equation}). Here we present a different method to get exact solutions of the radial equations employing hypergeometric functions. 

For the massive Dirac equation we make the separation of variables in the tetrad (\ref{csix:tetrad NP}), which is different from that used in Ref.\ \cite{Otchik:1985ih}; nevertheless we obtain the same system of differential equations for the radial functions.

We write the massive Dirac equation using the Newman-Penrose formalism in the form \cite{RS:Newman-1962-BWM-RS}, \cite{Chandrasekhar book}
\begin{eqnarray} \label{csix:Dirac-massive-NP}
(D + \epsilon - \rho)F_1 + (\delta^* + \tilde{\pi} - \alpha)F_2 = i \mu_* G_1, \nonumber \\
(\Delta + \tilde{\mu} - \gamma) F_2 + (\delta + \beta - \tau) F_1 = i \mu_* G_2, \\
(D + \epsilon^* - \rho^*) G_2 - (\delta + \tilde{\pi}^* - \alpha^*) G_1 = i \mu_* F_2, \nonumber \\
(\Delta + \tilde{\mu}^*-\gamma^*) G_1 - (\delta^* + \beta^*-\tau^*) G_2 = i \mu_* F_1, \nonumber
\end{eqnarray} 
where $M = \sqrt{2} \mu_*$ is the particle mass.

Employing the tetrad (\ref{csix:tetrad NP}) in $4D$ de Sitter metric and making the ansatz
\begin{align} \label{csix:ansatz-Dirac-massive}
&F_1 = {\rm e}^{-i \omega t} {}_{-\frac{1}{2}} Y_{jm}(\theta,\phi) \hat{F}_1(r),&  &F_2 = {\rm e}^{-i \omega t} {}_{\frac{1}{2}} Y_{jm}(\theta,\phi) \hat{F}_2(r) ,\nonumber \\
&G_1 = {\rm e}^{-i \omega t} {}_{-\frac{1}{2}} Y_{jm}(\theta,\phi) \hat{G}_1(r),& &G_2  = {\rm e}^{-i \omega t} {}_{\frac{1}{2}} Y_{jm}(\theta,\phi) \hat{G}_2(r),
\end{align}
where $\hat{F}_2=\hat{G}_1$ and $\hat{F}_1=\hat{G}_2$, it is possible to show that the massive Dirac equation in $4D$ de Sitter background simplifies to the system of differential equations for the radial functions \cite{Otchik:1985ih}, \cite{Chandrasekhar book}
\begin{eqnarray} \label{csix:radial-equations-FG}
\left(P\frac{{\rm d}}{{\rm d}r} - \frac{i \omega}{P} - \frac{r}{2 P l^2} + \frac{P}{r} \right) \hat{F}_1 + \left( \frac{j+\frac{1}{2}}{r} - i M \right) \hat{F}_2 = 0, \nonumber \\
\left(P\frac{{\rm d}}{{\rm d}r} + \frac{i \omega}{P} - \frac{r}{2 P l^2} + \frac{P}{r} \right) \hat{F}_2 + \left( \frac{j+\frac{1}{2}}{r} + i M \right) \hat{F}_1 = 0,
\end{eqnarray} 
where $j$ is a half-integer, $j \geq \frac{1}{2}$.

Making the changes $\hat{F}_1 \to R_1$, $\hat{F}_2 \to - R_2$, $r = z l$, $\tilde{\omega} = \omega l$, and $\tilde{M} = M l$ we find that the coupled differential equations (\ref{csix:radial-equations-FG}) become
\begin{align}  \label{csix:radial-equations-z}
\left( (1 -z^2)^{\frac{1}{2}} \frac{{\rm d}}{{\rm d}z} - \frac{i \tilde{\omega}}{(1 -z^2)^{\frac{1}{2}}}  - \frac{z}{2 (1 -z^2)^{\frac{1}{2}}}   \right. &  \left.  +  \frac{(1 -z^2)^{\frac{1}{2}}}{z}  \right)    R_1 \nonumber \\ 
&  = \left( \frac{j+\frac{1}{2}}{z} - i \tilde{M} \right) R_2 ,\nonumber 
\\
\left( (1 -z^2)^{\frac{1}{2}} \frac{{\rm d}}{{\rm d}z} + \frac{i \tilde{\omega}}{(1 -z^2)^{\frac{1}{2}}}  - \frac{z}{2 (1 -z^2)^{\frac{1}{2}}}  \right. & \left. + \frac{(1 -z^2)^{\frac{1}{2}}}{z} \right) R_2 \nonumber \\ 
&   = \left( \frac{j+\frac{1}{2}}{z} + i \tilde{M} \right) R_1.
\end{align} 

Proposing the ansatz
\begin{eqnarray} \label{csix:ansatz-R-1-R-2}
R_1 = h \tilde{R}_1, \nonumber \\
R_2 = h \tilde{R}_2,
\end{eqnarray} 
where 
\begin{equation}
h=\frac{1}{z(1-z^2)^{\frac{1}{4}}},
\end{equation} 
we can simplify the system of equations (\ref{csix:radial-equations-z}) to
\begin{eqnarray} \label{csix:radial-z-simplified}
\left((1-z^2)^{\frac{1}{2}}\frac{{\rm d}}{{\rm d}z} - \frac{i \tilde{\omega}}{(1-z^2)^{\frac{1}{2}}} \right)\tilde{R}_1 = \left(\frac{j + \frac{1}{2}}{z} - i \tilde{M}\right) \tilde{R}_2 , \nonumber \\
\left((1-z^2)^{\frac{1}{2}}\frac{{\rm d}}{{\rm d}z} + \frac{i \tilde{\omega}}{(1-z^2)^{\frac{1}{2}}} \right)\tilde{R}_2 = \left(\frac{j + \frac{1}{2}}{z} + i \tilde{M}\right) \tilde{R}_1.
\end{eqnarray} 

The system of differential equations (\ref{csix:radial-z-simplified}) is equivalent\footnote{Identifying $j+\tfrac{1}{2}$ with $m$ and $\tilde{M}$ with $\tilde{\mu}$.\label{footnote-ref}} to that given in Eq.\ (71) of Ref.\ \cite{Lopez-Ortega:2004cq} (from this system we can easily find Eqs.\ (4) of Ref.\ \cite{Otchik:1985ih}, as discussed in Ref.\ \cite{Lopez-Ortega:2004cq}). In Appendix \ref{appendix:solution-dirac-equation} we briefly describe the method used in Ref.\ \cite{Otchik:1985ih} to find exact solutions of the system of differential equations (\ref{csix:radial-z-simplified}) employing hypergeometric functions. In the following, we explain a different method to obtain exact solutions of the coupled equations (\ref{csix:radial-z-simplified}), which we believe is simpler than the one used in \cite{Otchik:1985ih}.

First, we make the ansatz (compare with that given in Eqs.\ (5) and (6) of Ref.\ \cite{Otchik:1985ih}, which we write in Eqs.\ (\ref{dsdS3:Dirac-mass-eq}) and (\ref{dS3:parameters-b_1-b_2}) of Appendix \ref{appendix:solution-dirac-equation})
\begin{eqnarray} \label{csix:ansatz-R-1-R2-f-1-f-2}
\tilde{R}_1 = (1-z^2)^{-\frac{1}{4}} (1-z)^{\frac{1}{2}} (f_1 - f_2),  \nonumber \\
\tilde{R}_2 = (1-z^2)^{-\frac{1}{4}} (1+z)^{\frac{1}{2}} (f_1 + f_2).
\end{eqnarray} 
Substituting these expressions for the functions $\tilde{R}_1$ and $\tilde{R}_2$ into the differential equations (\ref{csix:radial-z-simplified}) yield
\begin{align} \label{csix:radial-f1-f2}
(1-z^2)\frac{{\rm d}}{{\rm d}z} (f_1 + f_2) &+\left(\frac{1}{2} +   i \tilde{\omega} \right) (f_1 + f_2)  \nonumber \\ 
&- \frac{j + \frac{1}{2}}{z}(1-z) (f_1 - f_2) - i \tilde{M} (1 -z) (f_1 - f_2) = 0, \nonumber 
\end{align} 
\begin{align}
(1-z^2)\frac{{\rm d}}{{\rm d}z} (f_1 - f_2) &-\left(\frac{1}{2}  +  i \tilde{\omega} \right) (f_1 - f_2) \\ 
&- \frac{j + \frac{1}{2}}{z}(1+z) (f_1 + f_2) + i \tilde{M} (1 + z) (f_1 + f_2) = 0. \nonumber
\end{align} 

From previous equations, we find that the functions $f_1$ and $f_2$ must be solutions of the coupled differential equations
\begin{eqnarray} \label{csix:equations-f-1-f-2-mixed}
\left[ (1-z^2)\frac{{\rm d}}{{\rm d}z} - \left(\frac{j+\frac{1}{2}}{z} - i \tilde{M} z\right)  \right]f_1 = - \left[\frac{1}{2} + i \tilde{\omega} - (j + \frac{1}{2}) + i \tilde{M} \right] f_2 , \nonumber \\
\left[ (1-z^2)\frac{{\rm d}}{{\rm d}z} + \left(\frac{j+\frac{1}{2}}{z} - i \tilde{M}z \right)  \right]f_2 = - \left[\frac{1}{2} + i \tilde{\omega} + (j + \frac{1}{2}) - i \tilde{M}  \right] f_1.
\end{eqnarray} 
We easily find from this system of differential equations that the functions $f_1$ and $f_2$ satisfy 
\begin{align} \label{eq:sec4:f1-f2-coupled}
&\left\{(1-z^2)^2\frac{{\rm d}^2}{{\rm d}z^2} -2z(1-z^2)\frac{{\rm d}}{{\rm d}z} + (1-z^2)\left(\frac{j+\frac{1}{2}}{z^2} + i \tilde{M} \right) \right. \nonumber \\ 
&-\left. \left[ \left(\frac{1}{2} + i \tilde{\omega} \right)^2 - \left( j + \frac{1}{2} - i \tilde{M} \right)^2 \right]- \left(\frac{j+\frac{1}{2}}{z} - i \tilde{M} z \right)^2 \right\} f_1 = 0 , \\
&\left\{(1-z^2)^2\frac{{\rm d}^2}{{\rm d}z^2} -2z(1-z^2)\frac{{\rm d}}{{\rm d}z} - (1-z^2)\left(\frac{j+\frac{1}{2}}{z^2} + i \tilde{M} \right) \right. \nonumber \\ 
&-\left. \left[ \left(\frac{1}{2} + i \tilde{\omega} \right)^2 - \left( j + \frac{1}{2} - i \tilde{M} \right)^2 \right]- \left(\frac{j+\frac{1}{2}}{z} - i \tilde{M} z \right)^2 \right\} f_2 = 0 . \nonumber
\end{align}

Making in Eqs.\ (\ref{eq:sec4:f1-f2-coupled}) the change of variable $x =z^2$, and taking $f_1$ and $f_2$ equal to
\begin{eqnarray} \label{csix:f1-f2-ansatz}
f_1 = x^{C_1} (1-x)^{B_1} \hat{f}_1, \nonumber \\ 
f_2 = x^{C_2} (1-x)^{B_2} \hat{f}_2,
\end{eqnarray}  
where
\begin{align}
&C_1=\frac{\frac{1}{2}\pm j}{2}, &C_2&=\frac{\frac{1}{2}\pm (j+1)}{2},  \nonumber \\
&B_1 = \pm \frac{1}{2}\left(\frac{1}{2} + i \tilde{\omega} \right), &B_2& = \pm \frac{1}{2}\left(\frac{1}{2} + i \tilde{\omega} \right), 
\end{align} 
we find that the functions $\hat{f}_1$ and $\hat{f}_2$ are solutions of the differential equations
\begin{eqnarray}
\left[x (1-x)\frac{{\rm d}^2}{{\rm d}x^2} + [c_I - (a_I+b_I+1)x]\frac{{\rm d}}{{\rm d}x} - a_I b_I \right] \hat{f}_I = 0, 
\end{eqnarray} 
where $I=1,2,$ and the parameters $a_I,b_I$, and $c_I$ are equal to
\begin{align} \label{csix:f1-f2-constants}
&a_1=B_1 + C_1 + \frac{1}{2} + \frac{i \tilde{M}}{2} , &a_2&=B_2 + C_2 + \frac{1}{2} - \frac{i \tilde{M}}{2}   ,\nonumber \\
&b_1=B_1 + C_1 - \frac{i \tilde{M}}{2}  , &b_2&=B_2 + C_2 +  \frac{i \tilde{M}}{2}  , \nonumber \\
&c_1=2C_1+\frac{1}{2}, &c_2&=2C_2+\frac{1}{2}.
\end{align}
Therefore the functions $\hat{f}_1$ and $\hat{f}_2$ are solutions of hypergeometric type differential equations.\footnote{We point out that there are many papers in which the properties of the exact solutions of the massive Dirac equation in $4D$ de Sitter spacetime are investigated, see for example Refs.\ \cite{Du:2004jt}, \cite{Otchik:1985ih}, \cite{Cotaescu:1998ay}, \cite{Shishkin:1991ma}.}

As in previous sections, we have several sets of possible values for the parameters of the exact solutions depending on the values chosen for $B_1$, $B_2$, $C_1$, and $C_2$. (See previous sections and \cite{Abdalla:2002hg}, \cite{Lopez-Ortega:2005ep}.)

\subsection{Flux calculation}

The exact solutions of the massive Dirac equation previously found allow us to  study the absorption of a massive Dirac field by the cosmological horizon. As is well known,\footnote{Making appropriate changes in the notation.} Eq.\ (\ref{csix:Radial-component-J}) also determines the radial component of the conserved current for a massive Dirac field propagating in $4D$ de Sitter spacetime \cite{Chandrasekhar book}. Therefore, from Eq.\ (\ref{csix:Weyl-flux-equal-zero}), we must study the quantity $|R_1|^2 - |R_2|^2$ to calculate the radial flux of a massive Dirac field. Henceforth we analyze in detail the case $B_1=B_2=\tfrac{1}{2}( \tfrac{1}{2}+i \tilde{\omega})$, $C_1=\tfrac{1}{2}(j+\tfrac{1}{2})$, and $C_2 = C_1 + \tfrac{1}{2}$.

Employing Eqs.\ (\ref{csix:ansatz-R-1-R-2}), (\ref{csix:ansatz-R-1-R2-f-1-f-2}), and (\ref{csix:f1-f2-ansatz}) we can show that the functions $R_1$ and $R_2$ satisfy
\begin{equation} \label{csix:rest-of-R-1-R-2}
|R_1|^2 - |R_2|^2 = - \frac{2 z^{4C_1 - 1}}{ (1-z^2)^{\tfrac{1}{2}}} \{(\hat{f}_1 + \hat{f}_2)(\hat{f}_1^* + \hat{f}_2^*) - (1 -x)\hat{f}_2\hat{f}_2^*\}.
\end{equation} 
Thus it is necessary to analyze the terms in curly braces of Eq.\ (\ref{csix:rest-of-R-1-R-2}). For the case studied here, it is easy to show that the radial solution that is regular at the origin of the de Sitter metric is the function that includes
\begin{equation}\label{csix:solution-f-1}
\hat{f}_1 = {}_{2}F_{1}(a_1,b_1;c_1;x),
\end{equation} 
and for this case, from Eqs.\ (\ref{csix:equations-f-1-f-2-mixed}), we get that the function $\hat{f}_2$ is equal to
\begin{equation}\label{csix:solution-f-2}
\hat{f}_2 = - \frac{b_1}{c_1} {}_{2}F_{1}(a_2,b_2;c_2;x) = - \frac{b_1}{c_1} {}_{2}F_{1}(a_1,b_1+1;c_1+1;x),
\end{equation} 
which is regular at $r=0$.

Equations (\ref{csix:solution-f-1}) and (\ref{csix:solution-f-2}) yield
\begin{eqnarray} \label{csix:f-1-f-2-relations}
\hat{f}_1 + \hat{f}_2 = \frac{b_1^*}{c_1} {}_{2}F_{1} (a_1,b_1;c_1+1,x), \nonumber \\
\hat{f}_2 \hat{f}_2^* = \frac{|b_1|^2}{c_1^2} (1-x)^{-1} |{}_{2}F_{1} (a_1,b_1;c_1+1,x)|^2,
\end{eqnarray} 
hence
\begin{equation} \label{csix:f-1-f-2-condition-flux}
(\hat{f}_1 + \hat{f}_2)(\hat{f}_1^* + \hat{f}_2^*) - (1-x)\hat{f}_2\hat{f}_2^* = 0,
\end{equation} 
for all $r \in (0,l)$.

Taking into account Eqs.\ (\ref{csix:Weyl-flux-equal-zero}), (\ref{csix:rest-of-R-1-R-2}), (\ref{csix:f-1-f-2-relations}), and (\ref{csix:f-1-f-2-condition-flux}) we get that the radial flux of the massive Dirac field from or toward the cosmological horizon is zero, if we use the regular radial solutions at $r=0$ and the oscillation frequency $\omega$ is a real number. This result is similar to that found in the previous sections for the massless fields of spin $s=\tfrac{1}{2}, 1$ moving in $4D$ de Sitter and for a massless Dirac field propagating in $3D$ de Sitter, (see also Refs.\ \cite{Myung:2003cn}, \cite{Myung:2003ki}, \cite{Lopez-Ortega:2004cq}).

From the result that the cross section of the cosmological horizon is zero, in Ref.\ \cite{Myung:2003cn} was asserted that it does not emit Hawking radiation (see Introduction of \cite{Myung:2003cn}), an affirmation that contradicts the well established thermodynamical properties of de Sitter spacetime. We note that in Refs.\ \cite{dS3D-Lohiya}, \cite{Suzuki:1995nh}, \cite{Otchik:1985ih} a value different from zero was found for the absorption probability of de Sitter horizon, but in these references only radial functions that are divergent at $r=0$ were studied. We believe that the implications of the results of Refs.\ \cite{Myung:2003cn}, \cite{Myung:2003ki} and this paper on the emission properties of the de Sitter horizon must be investigated further.

An alternative is that our result follows from the assumption of thermal equilibrium made throughout this work and in Refs.\ \cite{Myung:2003cn}, \cite{Myung:2003ki}, \cite{Lopez-Ortega:2004cq}, (in our analysis the de Sitter cosmological horizon does not expands or shrinks).

\subsection{Quasinormal modes}

We can calculate the de Sitter QNMs of a massive Dirac field employing the exact solutions found in this section. When  $B_1=\tfrac{1}{2}( \tfrac{1}{2}+i \tilde{\omega})$ and $C_1=\tfrac{1}{2}(j+\tfrac{1}{2})$ from the property of the hypergeometric function given in Eq.\ (\ref{csix:hipergeometric-property-z-1-z}), we can write the radial function $R_1$ of Eq.\ (\ref{csix:ansatz-R-1-R-2}) in the form
\begin{align} \label{csix:Dirac-massive-1-x}
R_1&  = \frac{b_1}{c_1} \frac{(1-z)^{\tfrac{1}{2}} z^{j - \tfrac{1}{2}}}{(1-z^2)^{\tfrac{1}{2}}} \\ 
 & \times \left\{ \frac{\Gamma(c_1+1)}{\Gamma(c_1-b_1)} \left[ \frac{\Gamma(c_1-a_1-b_1)}{b_1 \Gamma(c_1-a_1)} {}_2F_1 (a_1,b_1;a_1+b_1+1-c_1;1-x) \right. \right. \nonumber  \\
&  \left. + x^{\tfrac{1}{2}} \frac{ \Gamma(c_1-a_1-b_1)}{\Gamma(c_1+1-a_1)}  {}_2F_1(a_1,b_1+1;a_1+b_1+1-c_1;1-x)\right](1-x)^{\tfrac{1}{2}\left(\tfrac{1}{2}+i \tilde{\omega}\right)}  \nonumber   
\end{align} 
\begin{align}
&  + \frac{\Gamma(c_1+1) \Gamma( a_1 +b_1 - c_1)}{\Gamma(a_1) \Gamma(b_1+1)}\left[ {}_2F_1(c_1-a_1,c_1-b_1; c_1 + 1 -a_1-b_1; 1 -x) \right. \nonumber \\
&\left. \left.+ x^{\tfrac{1}{2}} {}_2F_1(c_1-a_1+1, c_1-b_1; c_1 + 1 -a_1-b_1; 1 -x) \right](1-x)^{-\tfrac{1}{2}\left(\tfrac{1}{2}+i \tilde{\omega}\right)} \right\}.\nonumber
\end{align}

In Eq.\ (\ref{csix:Dirac-massive-1-x}), as $r \to l$, the term with the factor $(1-x)^{\tfrac{1}{2}\left(\tfrac{1}{2}+i \tilde{\omega}\right)}$ represents an ingoing wave, whereas the term with the factor $(1-x)^{-\tfrac{1}{2}\left(\tfrac{1}{2}+i \tilde{\omega}\right)}$ represents an outgoing wave. Therefore, in order to have purely outgoing oscillations as $r \to l$, it is necessary to satisfy the condition 
\begin{equation}
c_1 - a_1 = -n_1, \qquad \textrm{or}  \qquad c_1-b_1 = -n_1, \qquad n_1=0,1,2,\dots,
\end{equation}
which yield
\begin{equation} \label{csix:Dirac-massive-quasinormal-frequencies}
i \tilde{\omega} = j + 2 n_1 - i  \tilde{M} , \qquad \qquad i \tilde{\omega} = j + 2 n_1 + 1 + i  \tilde{M} ,
\end{equation} 
respectively.

We easy see that the factor $\Gamma(c_1+1)$ does not have poles, while the factor $\Gamma(c_1-a_1-b_1)$ has poles at
\begin{equation} \label{csix:poles-no-cancellation}
i \tilde{\omega} = n_1 - \frac{1}{2}.
\end{equation}  
From Eqs.\ (\ref{csix:Dirac-massive-quasinormal-frequencies}) and (\ref{csix:poles-no-cancellation}) we get that these values for $i \tilde{\omega}$ cannot be equal if $\tilde{M} \neq 0$, hence there is no pole cancellation. We note that for the frequencies (\ref{csix:Dirac-massive-quasinormal-frequencies}) the quantity $c_1-a_1-b_1$ is not an integer. 

We also find the values (\ref{csix:Dirac-massive-quasinormal-frequencies}) for $i \tilde{\omega}$ when we analyze the radial function $R_2$ by means of a similar method ($B_2=\tfrac{1}{2}( \tfrac{1}{2}+i \tilde{\omega})$ and $C_2=\tfrac{1}{2}(j+\tfrac{1}{2}) + \tfrac{1}{2}$ correspond to the case studied in detail).

An explicit calculation of the radial flux (\ref{csix:Weyl-flux-equal-zero}) shows that it is different from zero if the frequency of a massive Dirac field is equal to some of the frequencies (\ref{csix:Dirac-massive-quasinormal-frequencies}), and we note that for these frequencies the amplitude of the field decreases with time. Therefore the oscillations with frequencies equal to those given in Eq.\ (\ref{csix:Dirac-massive-quasinormal-frequencies}) are QNMs of a massive Dirac field moving in $4D$ de Sitter spacetime. 

If $M=0$, then the QNM frequencies (\ref{csix:Dirac-massive-quasinormal-frequencies}) are equal to those of Eq.\ (\ref{eq-quasinormal-frequencies-4D}) for $s=\tfrac{1}{2}$. Moreover, the frequencies (\ref{csix:Dirac-massive-quasinormal-frequencies}) coincide with those given in Eqs.\ (50) and (51) of Ref.\ \cite{Du:2004jt}, if the quantities $\kappa_+$ and $\kappa_-$ are positive and negatives integers, respectively. In Ref.\ \cite{Du:2004jt} the method used to calculate the QNM frequencies is different (and more complicate) to that applied in this section.

An analysis of the QNM frequencies (\ref{csix:Dirac-massive-quasinormal-frequencies}) show that, at least in four dimensions, the mass of a Dirac field does not have an inferior limit so that the QNM frequencies are well defined. As shown in Ref.\ \cite{Du:2004jt} (see also \cite{Abdalla:2002hg}), for a minimally coupled scalar field there is an inferior limit for its mass in order to have QNMs.

We stress that when the frequency of a Dirac field is equal to $i \tilde{\omega} = j - i\tilde{M} $ ($i \tilde{\omega} = j + 2 n_1 - i \tilde{M}$ when $n_1=0$), the method previously used to find the exact solutions of the massive Dirac equation is not applicable. Nevertheless, a detailed analysis of this case shows that the solutions found above hold replacing $i \tilde{\omega}$ by $j - i \tilde{M}$ in the  values of the different parameters of the solutions previously given. Moreover, if the frequency of a Dirac field is $i \tilde{\omega} = j  - i\tilde{M}$, an explicit calculation of the radial flux (\ref{csix:Weyl-flux-equal-zero}) prove that it is different from zero.

In the present section we find exact solutions of the Dirac equation in $4D$ de Sitter background, but taking into account the results of Ref.\ \cite{Lopez-Ortega:2004cq} we deduce that the exact solutions of the system of differential equations (\ref{csix:radial-equations-FG}) obtained in this section lead to exact solutions of the massive Dirac equation in $3D$ de Sitter metric, provided we make the identifications mentioned in footnote \ref{footnote-ref}. Thus, making these identifications, the frequencies (\ref{csix:Dirac-massive-quasinormal-frequencies}) are also QNM frequencies of a massive Dirac field propagating in $3D$ de Sitter spacetime. These values for the QNM frequencies are equal to those given in Eqs.\ (39) and (40) of Ref.\ \cite{Du:2004jt} if the quantity $\ell$ is a half-integer. To calculate these frequencies the authors of Ref.\ \cite{Du:2004jt} use a more complex method.

The method exploited in the present section also produces exact solutions of the massless Dirac equation when we take the limit $\mu \to 0$. A quick look shows that the solutions found in the present section are different from those studied in Section \ref{csix:dS4-massless-fields} of the present paper. We have not been able to find the relation between the two sets of exact solutions. 

We believe that the usefulness of the exact solutions previously found to study other physical phenomena occurring in de Sitter metric must be investigated further.

\section{Discussion}
\label{csix:End-comments}

In the present paper we studied the absorption and QNMs of some classical fields propagating in de Sitter spacetime. If the frequency $\omega$ is a real number and the fields have no divergences at the origin, the result found here and in Refs.\ \cite{Myung:2003cn}, \cite{Myung:2003ki}, \cite{Lopez-Ortega:2004cq} for the radial flux of a classical field moving in de Sitter background is the same (zero). We obtain this value for the radial flux studying the behavior of different fields, and we believe that the implications of this result must be analyzed further. An analogous fact is that the event horizon of an extreme BTZ black hole exhibits a vanishing absorption cross section, as shown in Refs.\ \cite{Gamboa:2000uc}.

Our results and those of Refs.\ \cite{Du:2004jt}, \cite{Abdalla:2002hg} suggest that for a massive classical field moving in de Sitter metric there are always QNMs. According to Ref.\ \cite{Natario:2004jd}, if the mass of the field is zero, then de Sitter QNMs exist only when the spacetime dimension is odd, but in Section \ref{csix:subsection:quasinormal-modes} (see also Appendix \ref{appendix:quasinormal-modes-4D}), we study an example with well defined QNM frequencies when the spacetime dimension is even and the fields are massless. Thus the affirmation of Ref.\ \cite{Natario:2004jd} on the existence of the de Sitter QNMs is not generally valid. Is it possible to get other massless fields with well defined QNM frequencies in de Sitter metric? (different from those found in Refs.\ \cite{Natario:2004jd}, \cite{Du:2004jt}, and the present paper). We believe that this point must be studied further.  

For massless fields the QNM frequencies found in this paper are purely imaginary as those found in Refs.\ \cite{Natario:2004jd}, \cite{Lopez-Ortega:2005ep}, \cite{Fernando:2003ai}, \cite{Lepe:2004kv}, \cite{Cardoso:2001bb}, and we notice that for the QNM frequencies calculated in the present paper the amplitude of the field diminishes with time.

We found that several parameters of the radial functions can take two values; as a consequence there are many possible combinations for the parameters depending on the values chosen. Nevertheless, some of our results indicate that the values chosen for the parameters do not change the physical conclusions obtained. (See also Refs.\ \cite{Abdalla:2002hg, Lopez-Ortega:2005ep}.)

An interesting question is to study the relevance of the results reported in the present paper for the dS-CFT correspondence \cite{Strominger:2001pn}, \cite{Abdalla:2002hg}, \cite{Abdalla:2002rm}. A natural extension of the present paper is to investigate if its results hold for the $D$-dimensional de Sitter spacetime. Finally, is there any relation between the purely imaginary de Sitter QNM frequencies and the algebraic special frequencies of the black holes?, (the algebraic special frequencies are purely imaginary \cite{csix:special-frequencies}).

\section{Acknowledgments}

I thank Dr.\ M.\ A.\ P\'erez Ang\'on for his interest in this paper, for valuable discussions, and also for proofreading the manuscript. I also thank Dr.\ G.\ F.\ Torres del Castillo and Dr.\ J.\ E.\ Rojas Marcial for their interest in this work and for valuable discussions. This work was supported by CONACyT and SNI (M\'exico).


\begin{appendix}

\section{Why $m$ is a half-integer?}
\label{appendix:half-integer-m}

In Section \ref{csix:dS3-Dirac-massless} we use the diagonal triad (\ref{triad}). As is well known, it is also possible to employ the Cartesian triad \cite{Brill:1957fx} for which the Gamma matrices are equal to
\begin{eqnarray} \label{appendix:gamma}
\gamma_t^{(c)} &=&  P  \gamma_1 , \nonumber \\
\gamma_r^{(c)} &=& \frac{1}{P}\left(\cos(\theta)\gamma_2 + \sin(\theta)\gamma_3 \right), \\
\gamma_\theta^{(c)} &=& r\left(- \sin(\theta)\gamma_2 + \cos(\theta)\gamma_3 \right),  \nonumber
\end{eqnarray} 
when the background is $3D$ de Sitter. In Eq.\ (\ref{gamma}), we previously define the matrices $\gamma_1$, $\gamma_2$, and $\gamma_3$ which appear in Eq.\ (\ref{appendix:gamma}). The relation between the solutions of the Dirac equation calculated in both triads is \cite{Brill:1957fx}
\begin{equation}
S \Psi = \Psi_{(c)},
\end{equation} 
where
\begin{equation} \label{appendix:matrix-forma}
S = \left( \begin{array}{cc} \cos(\theta/2) & -i \sin(\theta/2) \\  -i \sin(\theta/2) & \cos(\theta/2) 
\end{array} \right) = \exp\left(-\frac{i}{2} \theta \gamma_1\right).
\end{equation}

The solution $\Psi_{(c)}$ is single valued \cite{Brill:1957fx}, but we note that the matrix $S$ satisfies
\begin{equation}
S(\theta+2 \pi) = - S(\theta),
\end{equation} 
therefore the spinor $\Psi$ must satisfy the relation $\Psi(\theta + 2 \pi) = - \Psi(\theta)$. Since in Section \ref{csix:dS3-Dirac-massless} the $\theta$-dependence of the spinor $\Psi$ is of the form $e^{i m \theta}$, we find that the quantity $m$ takes the values $\pm \tfrac{1}{2}, \pm \tfrac{3}{2}, \pm \tfrac{5}{2},\dots$\footnote{The author gratefully acknowledges the many helpful suggestions of Dr.\ G.\ F.\ Torres del Castillo on the content of this Appendix.}

\section{Quasinormal modes in 4$D$ de Sitter spacetime}
\label{appendix:quasinormal-modes-4D}

In four dimensions, when the background spacetime is spherically symmetric, it is possible to show that the equations of motion for a scalar field, an electromagnetic field and an odd parity gravitational perturbation simplify to a Schr\"odinger type differential equation \cite{Chandrasekhar book}, \cite{Cardoso:2003sw}. For $4D$ de Sitter metric this equation is equal to
\begin{equation} \label{eq:app-Schodinger-eq}
\frac{{\rm d}^2 \Phi}{{\rm d} r_{*}^2} +[\omega^2 - V(r) ] \Phi = 0, 
\end{equation} 
where $r_{*}$ is the tortoise coordinate defined in Eq.\ (\ref{eq:sec2:tortoise}) and 
\begin{equation} \label{eq-app-effective-potential}
V(r) = P^2 \left[\frac{j(j+1)}{r^2}- \frac{2 \alpha}{l^2} \right],
\end{equation} 
with $\alpha=1$ for a massless minimally coupled scalar field and $\alpha=0$ for an electromagnetic field and an odd parity gravitational perturbation. For an even parity gravitational perturbation the effective potential $V(r)$ is a more complicated expression \cite{Cardoso:2003sw}, \cite{Cardoso:2001bb-II}.\footnote{Nevertheless see Ref.\ \cite{Mellor-Moss-1990}.} 

In de Sitter spacetime Eq.\ (\ref{eq:app-Schodinger-eq}) becomes
\begin{eqnarray} \label{eq-app-z-variable}
\left\{ (1-z^2)^2\frac{{\rm d}^2}{{\rm d}z^2} -2 z (1-z^2)\frac{{\rm d}}{{\rm d}z} + \tilde{\omega}^2 -(1-z^2)\left[\frac{j(j+1)}{z^2} - 2 \alpha \right] \right\} \Phi = 0,
\end{eqnarray} 
where $zl =r$ and $\tilde{\omega} = \omega l$ as previously. Making the change of variable $y=z^2$ \cite{Choudhury:2003wd}, \cite{Brady:1999wd}, \cite{Natario:2004jd}, \cite{Abdalla:2002hg} we find that Eq.\ (\ref{eq-app-z-variable}) takes the form
\begin{align} \label{eq-app-y-variable}
\left\{ \frac{{\rm d}^2}{{\rm d}y^2} + \left[\frac{1}{2y} - \frac{1}{1-y} \right]\frac{{\rm d}}{{\rm d}y}\right. + & \frac{\tilde{\omega}^2 -j(j+1) + 2 \alpha}{4y(1-y)}  \nonumber \\ 
&\left. \qquad \, + \frac{\tilde{\omega}^2}{4(1-y)^2} -\frac{j(j+1)}{4y^2}\right\} \Phi = 0 .
\end{align} 

Proposing a solution of Eq.\ (\ref{eq-app-y-variable}) in the form
\begin{equation}
\Phi = y^A (1-y)^B \tilde{\Phi},
\end{equation} 
we get that the function $\tilde{\Phi}$ satisfies a hypergeometric type differential equation
\begin{equation} \label{eq-app-hypergeometric}
\left\{ y (1-y)\frac{{\rm d}^2}{{\rm d}y^2} + [c -(a + b +1)y]\frac{{\rm d}}{{\rm d}y} -a b \right\} \tilde{\Phi} = 0,
\end{equation} 
where the parameters $a$, $b$, and $c$ are equal to
\begin{eqnarray} \label{eq-app-a-b-c-values}
a &=& A + B + \frac{1}{4} + \frac{1}{4}[1 + 8\alpha]^{\tfrac{1}{2}} ,\nonumber \\
b &=& A + B + \frac{1}{4} - \frac{1}{4}[1 + 8\alpha]^{\tfrac{1}{2}} , \\
c &=& 2 A + \frac{1}{2}, \nonumber
\end{eqnarray}
provided that
\begin{eqnarray}
A &= \left\{ \begin{array}{l} \frac{1}{2} + \frac{j}{2}, \\ - \frac{j}{2}, \end{array}\right. \qquad \qquad B= \pm \frac{i \tilde{\omega}}{2}.
\end{eqnarray}

In the following we study in detail the case $A=\tfrac{1}{2} + \tfrac{j}{2}$ and $B=\tfrac{i \tilde{\omega}}{2}$. Thus the solutions of the differential equation (\ref{eq-app-hypergeometric}) are \cite{b:DE-books} 
\begin{eqnarray}
\tilde{\Phi}^{(1)} &=& {}_{2}F_{1}(a,b;c;y),\\
\tilde{\Phi}^{(2)} &=& y^{1-c} {}_{2}F_{1}(a-c+1,b-c+1;2-c;y), \nonumber
\end{eqnarray} 
because the parameter $c$ is a half-integer.

Taking $\tilde{\Phi}^{(2)}$ as a solution of Eq.\ (\ref{eq-app-hypergeometric}), we find that the function $\Phi$ is divergent at the origin ($y=0$) for $j \geq 1$. In the present paper we are studying regular solutions at $r=0$, hence in the rest of this appendix we only consider $\tilde{\Phi}^{(1)}$. Thus
\begin{equation} \label{eq-app-physical-solution}
\Phi = y^{\tfrac{j}{2} + \tfrac{1}{2}} (1-y)^{\tfrac{i \tilde{\omega}}{2}} {}_{2}F_{1}(a,b;c;y).
\end{equation} 

If $c-a-b$ is not a integer, we can write Eq.\ (\ref{eq-app-physical-solution}) in the form \cite{b:DE-books} 
\begin{align} \label{eq-app-physical-field-1-y}
\Phi&= y^{\tfrac{j}{2} + \tfrac{1}{2}}\left\{\frac{\Gamma(c)\Gamma(c-a-b)}{\Gamma(c-a)\Gamma(c-b)}(1-y)^{\tfrac{i \tilde{\omega}}{2}} {}_{2}F_{1}(a,b;a+b-c+1;1-y)\right. \nonumber \\ 
&+ \left.\frac{\Gamma(c)\Gamma(a+b-c)}{\Gamma(a)\Gamma(b)}(1-y)^{-\tfrac{i \tilde{\omega}}{2}} {}_{2}F_{1}(c-a,c-b;c-a-b+1;1-y) \right\}.
\end{align} 
Since the field $\Phi$ depends on time as  $\textrm{e}^{-i \omega t}$, it follows that the first term in curly braces in Eq.\ (\ref{eq-app-physical-field-1-y}) represents an ingoing wave as $r \to l$ and the second term represent an outgoing wave as $r \to l$. As we are calculating the de Sitter QNM frequencies, the field must be purely outgoing as $r \to l$, that is, we must impose the condition
\begin{equation} \label{eq-app-c-a-condition}
c-a= -n_1, \quad \qquad \textrm{or} \quad \qquad c-b=-n_1.
\end{equation} 

Nevertheless, from the expressions for the parameters $a$, $b$, and $c$ given in Eq.\ (\ref{eq-app-a-b-c-values}), it is possible to show that if $c-a-b$ is not an integer, then $c-a$ and $c-b$ cannot be integers; therefore we cannot satisfy the conditions (\ref{eq-app-c-a-condition}).\footnote{We note that $(1+8\alpha)^{\tfrac{1}{2}}$ is an integer for $\alpha=0$ and $\alpha=1$. }

When $c-a-b$ is a negative integer,\footnote{If $c-a-b$ is zero or a positive integer, then there are no QNM frequencies (see footnote \ref{footnote-9}).} that is, $c-a-b=-n$, $n=1,2,3,\dots$ we can use the property (\ref{eq:hypergeometric-c-a-b-integer}) of the hypergeometric function  to write the field $\Phi$ as
\begin{align} \label{eq-app-c-a-b-integer}
\Phi &= y^{\tfrac{j}{2} + \tfrac{1}{2}} (1-y)^{\tfrac{i \tilde{\omega}}{2}} \left\{ \frac{\Gamma(a+b-n) \Gamma(n)}{\Gamma(a)\Gamma(b)} (1-y)^{-n} \right. \nonumber \\ 
&\times \sum_{p=0}^{n-1}\frac{(a-n)_p (b-n)_p}{p! (1-n)_p}(1-y)^p - \frac{(-1)^n \Gamma(a+b-n)}{\Gamma(a-n)\Gamma(b-n)} \sum_{p=0}^\infty \frac{(a)_p (b)_p}{p!(n+p)!}(1-y)^p \nonumber \\
& \left.  \times\frac{{}}{{}} [\textrm{In}(1-y)-\psi(p+1)-\psi(p+n+1)+\psi(a+p)+\psi(b+p)] \right\} .
\end{align}
The first term in curly braces of Eq.\ (\ref{eq-app-c-a-b-integer}) represents an outgoing wave and the second term represents an ingoing wave; therefore the field $\Phi$ is purely outgoing if we impose the condition
\begin{equation} \label{eq-app-conditions-quasinormal}
a- n = -n_1, \quad \qquad \textrm{or} \quad \qquad b - n = -n_1 .
\end{equation} 

From Eqs.\ (\ref{eq-app-conditions-quasinormal}) we find that the QNM frequencies are equal to
\begin{equation} \label{eq-app-quasinormal-freq}
i\tilde{\omega} = j + \frac{3}{2} + \frac{1}{2}(1+8\alpha)^{\tfrac{1}{2}} + 2 n_1, \qquad i\tilde{\omega} = j + \frac{3}{2} - \frac{1}{2}(1+8\alpha)^{\tfrac{1}{2}} + 2 n_1.
\end{equation} 
For an electromagnetic field and an axial gravitational perturbation these frequencies are equal to those calculated in Section \ref{csix:subsection:quasinormal-modes}, since for each $j$ the frequencies (\ref{eq-app-quasinormal-freq}) are even or odd integers, depending on the $j$ value. For a massless minimally coupled  scalar field our result is different from that of Refs.\ \cite{Choudhury:2003wd}, \cite{Brady:1999wd}.

In our analysis we do not include the mode $j=0$ of the massless scalar field. As shown in Ref.\ \cite{Brady:1999wd}, when this field is moving in de Sitter spacetime its modes $j=0$ and $j>0$ behave in a different way.

We can easily write the effective potential given in Eq.\ (\ref{eq-app-effective-potential}) in the form
\begin{equation} \label{eq-app-effective-potential-2}
V(r)=\frac{1}{l^2}\left\{ \frac{j(j+1)}{\sinh^2(r_*/l)}-\frac{2 \alpha}{\cosh^2(r_*/l)} \right\},
\end{equation} 
where $r=l\tanh(r_*/l)$ ($r_*$ is the tortoise coordinate). (See Refs.\ \cite{Du:2004jt} and \cite{Natario:2004jd} for the effective potentials of a massive scalar field and a gravitational perturbation, respectively, when these fields are propagating in $D$-dimensional de Sitter spacetime.)

In Ref.\ \cite{Beyer:1998nu} Beyer proved that for the P\"oschl-Teller potential
\begin{equation}
V_{PT}(x)=\frac{V_0}{\cosh^2(x)}, \qquad x \in {\mathbb{R}},
\end{equation} 
\textit{``after a large enough time $t_0$, the solutions of the wave equation corresponding to $C^{\infty}$ data with compact support can be expanded uniformly in time with respect to the quasinormal modes, thereby leading to absolutely convergent series''}  \cite{Beyer:1998nu}. For $r_* \geq 0$, is there a similar result for the potential (\ref{eq-app-effective-potential-2})? We believe that this point must be studied further.

\section{Otchik's method}
\label{appendix:solution-dirac-equation}

In Ref.\ \cite{Otchik:1985ih} Otchik proved that in order to find exact solutions of the massive Dirac equation in $4D$ de Sitter metric, it is necessary to solve the system of differential equations
\begin{eqnarray} \label{dS3d:Otchik equations}
\left[ (1- z^2)\frac{{\rm d}}{{\rm d}z} + i \tilde{\omega} \right] \tilde{R}_2 = (1- z^2)^{\frac{1}{2}} \left( \frac{\kappa}{z} - i \tilde{M} \right) \tilde{R}_1, \nonumber \\
\left[ (1- z^2)\frac{{\rm d}}{{\rm d}z} - i \tilde{\omega} \right] \tilde{R}_1 = (1- z^2)^{\frac{1}{2}} \left( \frac{\kappa}{z} + i \tilde{M} \right) \tilde{R}_2,
\end{eqnarray} 
where $z l = r$, $\tilde{\omega} = \omega l $, $\tilde{M} = M l$ as above and $\kappa =j+\tfrac{1}{2}$. This system of differential equations is equivalent to that given in Eq.\ (\ref{csix:radial-z-simplified}) (see Appendix of Ref.\ \cite{Lopez-Ortega:2004cq}).

To solve the system of differential equations (\ref{dS3d:Otchik equations}), Otchik proposes that the functions  $\tilde{R}_1$ and $\tilde{R}_2$ take the form
\begin{eqnarray} \label{dsdS3:Dirac-mass-eq}
\tilde{R}_1 &=& i[\tilde{b}_2 (1-\tilde{x})^{(\kappa+1)/2} \tilde{x}^{(1-i\tilde{\omega})/2} f_1 + \tilde{b}_1 (1-\tilde{x})^{\kappa/2} \tilde{x}^{-i\tilde{\omega}/2} f_2 ], \nonumber \\ 
\tilde{R}_2 &=& \tilde{b}_1 (1-\tilde{x})^{(\kappa+1)/2} \tilde{x}^{(1-i\tilde{\omega})/2} f_1 + \tilde{b}_2 (1-\tilde{x})^{\kappa/2} \tilde{x}^{-i\tilde{\omega}/2} f_2,    
\end{eqnarray}  
where $\tilde{x}=1-z^2$, and the functions $\tilde{b}_1$ and $\tilde{b}_2$ are equal to
\begin{eqnarray} \label{dS3:parameters-b_1-b_2}
\tilde{b}_1 &=& [\tilde{x}^{1/2} + i (1-\tilde{x})^{1/2}]^{1/2} + i [\tilde{x}^{1/2} - i (1-\tilde{x})^{1/2}]^{1/2}, \nonumber \\  
\tilde{b}_2 &=& - [\tilde{x}^{1/2} + i (1-\tilde{x})^{1/2}]^{1/2} + i [\tilde{x}^{1/2} - i (1-\tilde{x})^{1/2}]^{1/2}. 
\end{eqnarray}  

Substituting Eqs.\ (\ref{dsdS3:Dirac-mass-eq}) and (\ref{dS3:parameters-b_1-b_2}) into Eqs.\ (\ref{dS3d:Otchik equations}), we find that the functions $f_1$ and $f_2$ satisfy the coupled differential equations
\begin{eqnarray} \label{dS3:solutions-coupled-}
\tilde{x}(1-\tilde{x})\frac{{\rm d}f_1}{{\rm d} \tilde{x}} - [(i\tilde{\omega}-\tfrac{1}{2})(1-\tilde{x}) + (\kappa + \tfrac{1}{2})\tilde{x}]f_1 = i (-\kappa -\tfrac{1}{2}+i\tilde{\omega} + i\tilde{M} ) f_2, \nonumber \\  
\frac{{\rm d}f_2}{{\rm d} \tilde{x}} = i (\kappa -\tfrac{1}{2} - i\tilde{\omega} + i\tilde{M} ) f_1. \hspace{3.5cm}
\end{eqnarray} 
From Eqs.\ (\ref{dS3:solutions-coupled-}), after eliminating $f_1$, it is possible to show that the function $f_2$ must be a solution of the differential equation 
\begin{align}
\tilde{x}(1-\tilde{x})\frac{{\rm d}^2f_2}{{\rm d} \tilde{x}^2} &+ [(-i\tilde{\omega}+\tfrac{1}{2}) - (\kappa-i\tilde{\omega}+1)\tilde{x}]\frac{{\rm d}f_2}{{\rm d} \tilde{x}} \nonumber \\
&- \tfrac{1}{4} (\kappa-i\tilde{\omega}- i\tilde{M}+ \tfrac{1}{2})(\kappa-i\tilde{\omega} +i\tilde{M}-\tfrac{1}{2})f_2 = 0, 
\end{align} 
which is a hypergeometric type differential equation.

\end{appendix}

\end{document}